\documentclass[12pt]{article}

\usepackage{amsfonts}
\usepackage{amsmath}
\usepackage{amsthm}
\usepackage{cite}
\usepackage{epsfig}
\usepackage{latexsym}
\usepackage{paralist}
\usepackage{fancyhdr}
\usepackage{graphicx}
\numberwithin{equation}{section}
\usepackage[vcentermath]{youngtab}
\usepackage{young}
\usepackage{ytableau}
\usepackage{etex}
\usepackage{braket}
\usepackage{float}
\usepackage{extarrows}
\usepackage{tikz}
\usetikzlibrary{calc}
\usetikzlibrary{topaths}
\usetikzlibrary{decorations}
\usetikzlibrary{decorations.pathmorphing}
\usepackage{autobreak}

\usepackage{pict2e}   
\usepackage{animate}  

\usepackage{multirow}
\usepackage{bigdelim}
\usepackage{fancybox}
\usepackage{cals}

\def\be{\begin{equation}}
\def\ee{\end{equation}}
\def\bea{\begin{eqnarray}}
\def\eea{\end{eqnarray}}

\renewcommand{\thefootnote}{\fnsymbol{footnote}}

\begin{document}

\hfuzz=100pt
\title{{\Large \bf{On s-confinement in 3d $\mathcal{N}=2$ gauge theories\\with anti-symmetric tensors} }}
\date{}
\author{ Keita Nii$^a$\footnote{nii@itp.unibe.ch}
}
\date{\today}

\maketitle

\thispagestyle{fancy}
\cfoot{}
\renewcommand{\headrulewidth}{0.0pt}

\vspace*{-1cm}
\begin{center}
$^{a}${{\it Albert Einstein Center for Fundamental Physics }}
\\{{\it Institute for Theoretical Physics
}}
\\ {{\it University of Bern}}  
\\{{\it  Sidlerstrasse 5, CH-3012 Bern, Switzerland}}

\end{center}

\begin{abstract}
We elaborate on s-confinement phases in three-dimensional $\mathcal{N}=2$ supersymmetric gauge theory, especially focusing on the $SU(N)$ and $USp(2N)$ gauge theories with anti-symmetric tensors and (anti-)fundamental matters. This will elucidate a quantum structure of the Coulomb moduli space of vacua. We stress the importance of so-called dressed Coulomb branch operators for describing these s-confinement phases. The 3d s-confinement phases are highly richer than the 4d ones since there is no chiral anomaly constraint on the matter contents.    
\end{abstract}

\renewcommand{\thefootnote}{\arabic{footnote}}
\setcounter{footnote}{0}

\newpage
\tableofcontents 


\section{Introduction}
Supersymmetric gauge theories can exhibit various low-energy phases depending on gauge groups, matter representations, spacetime dimensions and so on. By tuning the matter content, we can observe, for instance, SUSY breaking, quantum-deformed moduli space, s-confinement, non-abelian Coulomb phases and so on \cite{Seiberg:1994bz, Seiberg:1994pq}. Among these phases, the s-confinement phase is very useful since the dual description does not include any gauge interaction and there are only gauge-singlet chiral superfields with some confining superpotential. It is easy to calculate various low-energy quantities by using the dual confining theory. The s-confinement phase in 4d $\mathcal{N}=1$ supersymmetric gauge theories was found in \cite{Seiberg:1994bz} and then classified in \cite{Csaki:1996zb}. The 3d s-confinement phase was studied, for example, in \cite{Aharony:1997bx, Karch:1997ux, Aharony:1997gp, Aharony:2011ci, Csaki:2014cwa, Amariti:2015kha, Nii:2016jzi, Nii:2018erm, Nii:2018tnd, Nii:2018wwj}.

In this paper, we will investigate the 3d s-confinement phases in 3d $\mathcal{N}=2$ $SU(N)$ and $USp(2N)$ gauge theories with anti-symmetric matters. Compared to the 4d s-confinement, the 3d s-confinement phases are very rich since the 3d theory can have ``chiral'' matter contents where the corresponding 4d theories include a chiral gauge anomaly and are ill-defined. For those ``chiral'' theories, the Coulomb branch operator is not gauge-invariant and seems to be lifted from the moduli space. However, we can define the so-called dressed Coulomb branch (monopole) operators which parametrize the quantum Coulomb branch directions \cite{Csaki:2014cwa, Amariti:2015kha}. In this paper, we classify the 3d s-confinement phases with anti-symmetric tensors and find that these theories are related via various deformation to each other and to the 4d s-confinement phases. As a consistency check of our analysis, we will compute the superconformal indices by using the electric and magnetic (confinement) descriptions for the $SU(4)$ and $USp(4)$ cases and will find a perfect agreement.  

The rest of this paper is organized as follows.
In Section 2 and 3, we study the s-confinement phases in 3d $\mathcal{N}=2$ $SU(2N)$ and $SU(2N+1)$ gauge theories, respectively. These theories include one or two anti-symmetric matters.
In Section 4 and 5, we investigate the s-confinement phases in 3d $\mathcal{N}=2$ $SU(4)$ and $SU(5)$ gauge theories. Some examples will include three anti-symmetric matters. 
In Section 6, we will study the s-confinement phases in 3d $\mathcal{N}=2$ $USp(2N)$ gauge theories with anti-symmetric matters. 
In Section 7, we summarize our findings and discuss future directions.

\section{$SU(2N)$ gauge theories}
In this section, we study the s-confinement phases in the 3d $\mathcal{N}=2$ $SU(2N)$ gauge theories with anti-symmetric tensors and (anti-)fundamental matters. We first discuss a generic structure of the Coulomb branch \cite{Csaki:2014cwa, Amariti:2015kha, Nii:2016jzi} which can be quantum-mechanically massless and stable. The bare Coulomb branch operator denoted by $Y_{SU(2N-2)}^{bare}$ induces the following gauge symmetry breaking
\begin{align}
SU(2N) &\rightarrow SU(2N-2) \times U(1)_1 \times U(1)_2 \nonumber \\
{\tiny \yng(1,1)} &\rightarrow {\tiny \yng(1,1)}_{\,0,-2}+{\tiny \yng(1)}_{\,1,N-2}+{\tiny \yng(1)}_{-1,N-2} +1_{0,2(N-1)}  \\
{\tiny \overline{\yng(1,1)}} &\rightarrow {\tiny \overline{\yng(1,1)}}_{\,0,2}+{\tiny \overline{\yng(1)}}_{\,-1,-(N-2)}+{\tiny \overline{\yng(1)}}_{1,-(N-2)} +1_{0,-2(N-1)}  \\
{\tiny \yng(1)} & \rightarrow {\tiny \yng(1)}_{\, 0,-1}+1_{1,N-1}+1_{-1,N-1}   \\
{\tiny \overline{ \yng(1)}}  &\rightarrow {\tiny \overline{ \yng(1)}}_{\, 0,1} +1_{-1,-(N-1)} +1_{1,-(N-1)},
\end{align}
where the Coulomb branch corresponds to the first $U(1)_1$ generator. $Y_{SU(2N-2)}^{bare}$ is constructed by dualizing the $U(1)_1$ vector superfield. The components charged under the $U(1)_1$ symmetry are all massive and integrated out. Notice that the anti-symmetric matter reduces to the two massless components ${\tiny \yng(1,1)}_{\,0,-2}$ and $1_{0,2(N-1)}$. This fact leads to a very rich structure of the dressed Coulomb brach as we will see in the following subsections. 

When the theory is ``chiral,'' the bare Coulomb branch cannot be gauge-invariant. Suppose that the 3d $\mathcal{N}=2$ $SU(2N)$ gauge theory includes $F$ fundamental, $\bar{F}$ anti-fundamental, $F_A$ anti-symmetric and $\bar{F}_A$ anti-symmetric-bar matters. In this set-up, the mixed Chern-Simons term between the $U(1)_1$ and $U(1)_2$ symmetries is generated as 
\begin{align}
k_{eff}^{U(1)_1,U(1)_2} =(2N-2)(N-2) (F_A-\bar{F}_A) +(N-1)(F-\bar{F}).
\end{align}
Therefore, the bare Coulomb branch $Y_{SU(2N-2)}^{bare}$ has a $U(1)_2$ charge $-k_{eff}^{U(1)_1,U(1)_2}$ \cite{Intriligator:2013lca}. Notice that, for the theories with special matter contents such that for the corresponding 4d theories have no gauge anomaly, the $U(1)_2$ charge of $Y_{SU(2N-2)}^{bare}$ is canceled. 
In order to parametrize the Coulomb branch, we have to define gauge-invariant operators by dressing the bare Coulomb branch with the massless matter fields. Depending on the value of $(F,\bar{F},F_A,\bar{F}_A)$, the precise forms of the dressed operators will change. In the following subsections, we give a list of s-confinement by defining the dressed Coulomb branch operators.

\subsection{$SU(2N)$ with $2 \, \protect\Young[-0.5]{11}+4 \, \protect\Young[0]{1}$}
The ultra-violet (UV) description is a 3d $\mathcal{N}=2$ $SU(2N)$ gauge theory with two anti-symmetric tensors and four fundamental matters. The matter fields and their quantum numbers are summarized in Table \ref{SU(2N)2anti40}. Since the theory is ``chiral'' in a sense that the corresponding 4d theory has a chiral gauge anomaly, the bare Coulomb branch operator $Y_{SU(2N-2)}^{bare}$ is not gauge-invariant. The dressed operator is defined by
\begin{align}
Y^{d}:=Y_{SU(2N-2)}^{bare} (1_{0,2(N-1)})^{2N-2}\sim Y_{SU(2N-2)}^{bare} A^{2N-2},
\end{align}
where the flavor indices of $A^{2N-2}$ should be totally symmetrized. 
The low-energy dynamics is described by the gauge invariant fields listed in Table \ref{SU(2N)2anti40} and a confining superpotential
\begin{align}
W= Y^d \left( T_N T_{N-2} +T_{N-1}^2 \right),
\end{align}
which is consistent with all the symmetries of Table \ref{SU(2N)2anti40}. The case for $N=2$ will be individually discussed in Section 4 and its superconformal indices will be tested.

\begin{table}[H]\caption{3d $\mathcal{N}=2$ $SU(2N)$ with $2\, {\tiny \protect\yng(1,1)} +4\, {\tiny \protect\yng(1)}$} 
\begin{center}
\scalebox{0.98}{
  \begin{tabular}{|c||c||c|c|c|c|c| } \hline
  &$SU(2N)$&$SU(2)$&$SU(4)$&$U(1)$&$U(1)$&$U(1)_R$  \\ \hline
 $A$ &${\tiny \yng(1,1)}$& ${\tiny \yng(1)}$ &1&1&0&0 \\
 $Q$ & ${\tiny \yng(1)}$ &1& ${\tiny \yng(1)}$&0&1&0 \\  \hline
 $T_N:=A^N$&1&\scriptsize $N$-th symm.&1&$N$&0&0  \\
 $T_{N-1}:=A^{N-1}Q^2$&1&\scriptsize $(N-1)$-th symm.&${\tiny \yng(1,1)}$&$N-1$&2&0  \\
 $T_{N-2}:=A^{N-2}Q^4$&1& \scriptsize $(N-2)$-th symm.&1&$N-2$&4&0  \\  \hline
$Y_{SU(2N-2)}^{bare}$ &\tiny $U(1)_2:$ $-4(N-1)^2$&1&1&$4-4N$&$-4$&2  \\
 \small  $Y^{d}:=Y_{SU(2N-2)}^{bare} A^{2N-2}$ &1&\scriptsize $(2N-2)$-th symm.&1&$2-2N$&$-4$&2  \\
 \hline
  \end{tabular}}
  \end{center}\label{SU(2N)2anti40}
\end{table}

\subsection{$SU(2N)$ with $2 \, \protect\Young[-0.5]{11}+3 \, \protect\Young[0]{1}+  \overline{\protect\Young[0]{1}}$}
The next example is a 3d $\mathcal{N}=2$ $SU(2N)$ gauge theory with two anti-symmetric tensors, three fundamental matters and one anti-fundamental matter. Table \ref{SU(2N)2anti31} summarizes the quantum numbers of the elementary fields and the moduli coordinates. The bare Coulomb branch operator $Y_{SU(2N-2)}^{bare}$ has a non-zero $U(1)_2$ charge $-2(N-1)(2N-3)$ and the dressed operator is defined by 
\begin{align}
Y^d:=Y_{SU(2N-2)}^{bare} (1_{0,2(N-1)})^{2N-3} \sim Y_{SU(2N-2)}^{bare} A^{2N-3},
\end{align}
where the flavor indices of $A^{2N-3}$ are symmetrized. 
The low-energy effective theory is described by the gauge-invariant chiral superfields defined in Table \ref{SU(2N)2anti31} and a confining superpotential
\begin{align}
W= T^d \left( MT_{N-1} T_N +T_N P_3 +T_{N-1} P_1 \right),
\end{align}
which is consistent with all the symmetries. The case with $N=2$ will be discussed also in Section 4. 

By introducing a rank-one vev to $M$, the theory flows to a 3d $\mathcal{N}=2$ $SU(2N-1)$ gauge theory with two anti-symmetric tensors and four fundamental matters, which is again s-confining and will be discussed in Section 3. 

\begin{table}[H]\caption{3d $\mathcal{N}=2$ $SU(2N)$ with $2\, {\tiny \protect\yng(1,1)} +3\, {\tiny \protect\yng(1)}+ {\tiny \overline{\protect\yng(1)}}$} 
\begin{center}
\scalebox{0.88}{
  \begin{tabular}{|c||c||c|c|c|c|c|c| } \hline
  &$SU(2N)$&$SU(2)$&$SU(3)$&$U(1)$&$U(1)$&$U(1)$&$U(1)_R$  \\ \hline
 $A$ &${\tiny \yng(1,1)}$& ${\tiny \yng(1)}$ &1&1&0&0&0 \\
 $Q$ & ${\tiny \yng(1)}$ &1& ${\tiny \yng(1)}$&0&1&0&0 \\  
$\tilde{Q}$ &${\tiny \overline{ \yng(1)}}$&1&1&0&0&1&0  \\  \hline
$M:=Q \tilde{Q}$ &1&1&${\tiny \yng(1)}$ &0&1&1&0  \\
$T_N:=A^N$ &1&\scriptsize$N$-th symm.&1&$N$&0&0&0  \\
 $T_{N-1}:=A^{N-1}Q^2$ &1&\scriptsize$(N-1)$-th symm.&${\tiny \overline{ \yng(1)}}$&$N-1$&2&0&0  \\
 $P_1:=A^{N-1}(A\tilde{Q})Q$  &1&\scriptsize$(N-2)$-th symm.&${\tiny \yng(1)}$ &$N$&1&1&0  \\
$P_3:=A^{N-2}(A \tilde{Q})Q^3$ &1&\scriptsize$(N-3)$-th symm.&1&$N-1$&3&1&0  \\  \hline
  $Y_{SU(2N-2)}^{bare}$&\tiny \begin{tabular}{c}$U(1)_2$ charge: \\ $-2(N-1)(2N-3)$　\end{tabular}&1&1&$4-4N$&$-3$&$-1$&2 \\
   $Y^d:=Y_{SU(2N-2)}^{bare} A^{2N-3} $&1&\scriptsize$(2N-3)$-th symm.&1&$1-2N$&$-3$&$-1$&2  \\  \hline
  \end{tabular}}
  \end{center}\label{SU(2N)2anti31}
\end{table}

\subsection{$SU(2N)$ with $2  \, \protect\Young[-0.5]{11}+2\, \protect\Young[0]{1}+2\, \overline{\protect\Young[0]{1}} $}
The UV description is a 3d $\mathcal{N}=2$ $SU(2N)$ gauge theory with two anti-symmetric tensors, two fundamental matters and two anti-fundamental matters. The $U(1)_2$ charge of the bare Coulomb branch $Y^{bare}_{SU(2N-2)}$ is $-2(N-1)(2N-4)$. Notice that the case with $N=2$ is ``vector-like'' and $Y^{bare}_{SU(2N-2)}$ is gauge-invariant \cite{Csaki:2014cwa, Nii:2016jzi}. 
The dressed (gauge-invariant) operator is defined by
\begin{align}
Y^d:=Y_{SU(2N-2)}^{bare} (1_{0,2(N-1)})^{2N-4}   \sim Y_{SU(2N-2)}^{bare} A^{2N-4},
\end{align}
where the flavor indices of $A^{2N-4}$ are totally symmetrized.
Table \ref{SU(2N)2anti22} summarizes the quantum numbers of the elementary fields and the moduli coordinates. The low-energy dynamics is described by the moduli fields in Table \ref{SU(2N)2anti22} and the superpotential 
\begin{align}
W=Y^d \left( M^2T_N^2+MP_1 T_N +T_NR+T_{N-1}P_2+ P_1^2  + \bar{B}T_{N-1} T_N\right).
\end{align}
The case with $N=2$ will be independently discussed in Section 4 where we will test the superconformal indices by using the electric and magnetic descriptions. 

By introducing a rank-one vev to $M$, the theory flows to a 3d $\mathcal{N}=2$ $SU(2N-1)$ gauge theory with two anti-symmetric tensors, three fundamental matters and one anti-fundamental matter, which will exhibit s-confinement and will be discussed in Section 3. 

\begin{table}[H]\caption{3d $\mathcal{N}=2$ $SU(2N)$ with $2\, {\tiny \protect\yng(1,1)} +2\, {\tiny \protect\yng(1)}+2 \, {\tiny \overline{\protect\yng(1)}}$} 
\begin{center}
\scalebox{0.81}{
  \begin{tabular}{|c||c||c|c|c|c|c|c|c| } \hline
  &$SU(2N)$&$SU(2)$&$SU(2)$&$SU(2)$&$U(1)$&$U(1)$&$U(1)$&$U(1)_R$  \\ \hline
 $A$ &${\tiny \yng(1,1)}$& ${\tiny \yng(1)}$ &1&1&1&0&0&0 \\
 $Q$ & ${\tiny \yng(1)}$ &1& ${\tiny \yng(1)}$&1&0&1&0&0 \\  
$\tilde{Q}$ &${\tiny \overline{ \yng(1)}}$&1&1&${\tiny \yng(1)}$&0&0&1&0  \\  \hline
$M:=Q \tilde{Q}$ &1&1&${\tiny \yng(1)}$&${\tiny \yng(1)}$ &0&1&1&0  \\
$\bar{B}:=A \tilde{Q}^2$ &1& ${\tiny \yng(1)}$&1&1&1&0&2&0  \\
$T_N:=A^N$ &1&\scriptsize$N$-th symm.&1&1&$N$&0&0&0  \\
 $T_{N-1}:=A^{N-1}Q^2$ &1&\scriptsize$(N-1)$-th symm.&1&1&$N-1$&2&0&0  \\
 $P_1:=A^{N-1}(A \tilde{Q})Q$&1&\scriptsize$(N-2)$-th symm.&${\tiny \yng(1)}$&${\tiny \yng(1)}$&$N$&1&1&0 \\
$P_2:=A^{N-1} (A\tilde{Q})^2$ &1&\scriptsize$(N-3)$-th symm.&1&1&$N+1$&0&2&0 \\ 
 $R:= A^{N-2}(A \tilde{Q})^2 Q^2$&1&\scriptsize$(N-4)$-th symm.&1&1&$N$&2&2&0 \\  \hline
 $Y_{SU(2N-2)}^{bare}$ &\tiny \begin{tabular}{c}$U(1)_2$ charge: \\ $-2(N-1)(2N-4)$　\end{tabular}&1&1&1&$4-4N$&$-2$&$-2$&2 \\ 
\small  $Y^d:=Y_{SU(2N-2)}^{bare} A^{2N-4} $&1&\scriptsize$(2N-4)$-th symm.&1&1&$-2N$&$-2$&$-2$&2 \\  \hline
  \end{tabular}}
  \end{center}\label{SU(2N)2anti22}
\end{table}

\subsection{$SU(2N)$ with $2 \, \protect\Young[-0.5]{11}+ \protect\Young[0]{1}+3 \, \overline{\protect\Young[0]{1}}$}
The UV description is a 3d $\mathcal{N}=2$ $SU(2N)$ gauge theory with two anti-symmetric tensors, one fundamental matter and three anti-fundamental matters. Table \ref{SU(2N)2anti13} summarizes the quantum numbers of the elementary degrees of freedom and the moduli operators. Since the theory is ``chiral,'' the bare Coulomb branch operator $Y_{SU(2N-2)}^{bare}$ must be dressed by the matter chiral superfields. The gauge invariant combination becomes
\begin{align}
Y^d:= Y_{SU(2N-2)}^{bare}   (1_{0,2(N-1)})^{2N-5}  \sim Y_{SU(2N-2)}^{bare} A^{2N-5}.
\end{align}
The low-energy dynamics is described by the gauge invariant composites listed in Table \ref{SU(2N)2anti13} with a  confining superpotential
\begin{align}
W= Y^d \left(T_N \bar{B}P_1+  P_1P_2 +T_NP_3 +T_N P_2M\right).
\end{align}
The case with $N=2$ will be discussed in Section 4 where we will find that the electric and confinement descriptions show an identical superconformal indices. By introducing a rank-one vev to $M$, the theory flows to a 3d $\mathcal{N}=2$ $SU(2N-1)$ gauge theory with two anti-symmetric tensors, two (anti-)fundamental flavors, which will be discussed in Section 3 and show s-confinement. 

For $N=3$,  we need a special treatment since the Higgs branch operator $P_3$ is not available and since there is an additional Coulomb branch operator. Let us consider the dressed operator 
\begin{align}
\left. Y_{SU(2N-2)}^{bare}   (1_{0,2(N-1)})^{2N-5}  1_{0,2(N-1)}  \left( {\tiny \yng(1,1)}_{\,0,-2} \right)^{N-1} \right|_{N=3}, \label{dressdress}
\end{align}
where the flavor indices have nine components. The eight components of them are identified with the product $Y^d   T_N |_{N=3}$ while the remaining one should be regarded as an additional Coulomb branch operator $Y_{SU(2N-2)}^{dressed, A}$ whose flavor indices are all contracted. For higher $N$, all the components of \eqref{dressdress} are identified with $Y^d   T_N $ and those additional operators are not necessary. 

\begin{table}[H]\caption{3d $\mathcal{N}=2$ $SU(2N)$ with $2\, {\tiny \protect\yng(1,1)} + {\tiny \protect\yng(1)}+3\, {\tiny \overline{\protect\yng(1)}}$} 
\begin{center}
\scalebox{0.86}{
  \begin{tabular}{|c||c||c|c|c|c|c|c| } \hline
  &$SU(2N)$&$SU(2)$&$SU(3)$&$U(1)$&$U(1)$&$U(1)$&$U(1)_R$  \\ \hline
 $A$ &${\tiny \yng(1,1)}$& ${\tiny \yng(1)}$ &1&1&0&0&0 \\
 $Q$ & ${\tiny \yng(1)}$ &1&1&0&1&0&0 \\  
$\tilde{Q}$ &${\tiny \overline{ \yng(1)}}$&1& ${\tiny \yng(1)}$&0&0&1&0  \\  \hline
$M:=Q \tilde{Q}$ &1&1&${\tiny \yng(1)}$ &0&1&1&0  \\
$\bar{B}:=A \tilde{Q}^2$&1&${\tiny \yng(1)}$&${\tiny \overline{\yng(1)}}$&1&0&2&0  \\
$T_N:=A^N$ &1&\scriptsize$N$-th symm.&1&$N$&0&0&0  \\
 $P_1:=A^{N-1}(A\tilde{Q})Q$  &1&\scriptsize$(N-2)$-th symm.&${\tiny \yng(1)}$ &$N$&1&1&0  \\
  $P_2:=A^{N-1}(A\tilde{Q})^2$  &1&\scriptsize$(N-3)$-th symm.&${\tiny \overline{ \yng(1)}}$ &$N+1$&0&2&0  \\
$P_3:=A^{N-2}(A \tilde{Q})^3Q$ &1&\scriptsize$(N-5)$-th symm.&1&$N+1$&1&3&0  \\  \hline
  $Y_{SU(2N-2)}^{bare}$&\tiny \begin{tabular}{c}$U(1)_2$ charge: \\ $-2(N-1)(2N-5)$　\end{tabular}&1&1&$4-4N$&$-1$&$-3$&2 \\
   $Y^d:=Y_{SU(2N-2)}^{bare} A^{2N-5} $&1&\scriptsize$(2N-5)$-th symm.&1&$-2N-1$&$-1$&$-3$&2  \\  \hline
  \end{tabular}}
  \end{center}\label{SU(2N)2anti13}
\end{table}

\subsection{$SU(2N)$ with $2 \, \protect\Young[-0.5]{11}+4 \, \overline{\protect\Young[0]{1}}$}
The next example is a 3d $\mathcal{N}=2$ $SU(2N)$ gauge theory with two anti-symmetric matters and four anti-fundamental matters. The $U(1)_2$ charge of the bare Coulomb branch $Y_{SU(2N-2)}^{bare}$ is $-2(N-1)(2N-6)$ and the dressed operator is defined by
\begin{align}
Y^{d}:=Y_{SU(2N-2)}^{bare}    (1_{0,2(N-1)})^{2N-6}  \sim Y_{SU(2N-2)}^{bare} A^{2N-6},
\end{align}
where the flavor indices of $A$ should be symmetrized. 
For $N=2$, $Y_{SU(2N-2)}^{bare}$ is positively charged and it should be instead dressed by ${\tiny \yng(1,1)}_{\,0,-2} \in A$ and this case will be studied in Section 4. Table \ref{SU(2N)2anti04} summarizes the quantum numbers of the matter content and the moduli coordinates. The low-energy effective theory is described by the superpotential 
\begin{align}
W= Y^d  \left( T_N T_{N-2}+ T_{N-1}^2 +T_N^2 \bar{B}^2 + \bar{B}T_{N-1} T_N \right).
\end{align}

For $N=3 $ and $4$, where the Higgs branch operator $T_{N-2}$ is not available, we need a special care about the dressed Coulomb branch since there are additional Coulomb branches. For $N=3$, the bare Coulomb branch $Y_{SU(2N-2)}^{bare}$ is gauge-invariant and can be used as a moduli coordinate. In addition to this bare operator, we can also define the dressed operator
\begin{align}
Y_{SU(2N-2)}^{dressed, A} := \left. Y_{SU(2N-2)}^{bare} 1_{0,2(N-1)} \left( {\tiny \yng(1,1)}_{\,0,-2} \right)^{N-1} \right|_{N=3},
\end{align}
where the right-hand side has six components from the flavor indices. The four components are identified with the operator product $Y^d   T_N |_{N=3}$ and the remaining two should be regarded as an independent dressed Coulomb branch operator $Y_{SU(2N-2)}^{dressed, A}$ which is fundamental under the $SU(2)$ global symmetry. For $N=4$, we need to introduce an additional dressed operator
\begin{align}
Y_{SU(2N-2)}^{dressed, A} := \left. Y_{SU(2N-2)}^{bare} (1_{0,2(N-1)})^{2N-6} 1_{0,2(N-1)} \left( {\tiny \yng(1,1)}_{\,0,-2} \right)^{N-1} \right|_{N=4},
\end{align}
where all the flavor indices of $A$'s are contracted and $Y_{SU(2N-2)}^{dressed, A}$ becomes a singlet. Notice that $Y^d   T_N |_{N=4}$ has 15 components while $ \left.(1_{0,2(N-1)})^{2N-6} 1_{0,2(N-1)} \left( {\tiny \yng(1,1)}_{\,0,-2} \right)^{N-1} \right|_{N=4}$ has 16 components and that this difference can be explained by a new dressed operator $Y_{SU(2N-2)}^{dressed, A} $. For more higher $N$, there is no need to introduce additional operators dressed by $1_{0,2(N-1)} \left( {\tiny \yng(1,1)}_{\,0,-2} \right)^{N-1}$.

\begin{table}[H]\caption{3d $\mathcal{N}=2$ $SU(2N)$ with $2\, {\tiny \protect\yng(1,1)} +4\, {\tiny \overline{\protect\yng(1)}}$} 
\begin{center}
\scalebox{0.9}{
  \begin{tabular}{|c||c||c|c|c|c|c| } \hline
  &$SU(2N)$&$SU(2)$&$SU(4)$&$U(1)$&$U(1)$&$U(1)_R$  \\ \hline
 $A$ &${\tiny \yng(1,1)}$& ${\tiny \yng(1)}$ &1&1&0&0 \\
 $\tilde{Q}$ & ${\tiny \overline{ \yng(1)}}$ &1& ${\tiny \yng(1)}$&0&1&0 \\  \hline
$\bar{B}:=A\tilde{Q}^2 $&1&${\tiny \yng(1)}$&${\tiny \yng(1,1)}$&1&2&0 \\
$T_N:=A^N$&1&\scriptsize $N$-th symm.&1&$N$&0&0  \\
$T_{N-1}:=A^{N-1}(A \tilde{Q})^2$&1&\scriptsize $(N-3)$-th symm.&${\tiny \yng(1,1)}$&$N+1$&2&0  \\
$T_{N-2}:=A^{N-2}(A \tilde{Q})^4$&1&\scriptsize $(N-6)$-th symm.&1&$N+2$&4&0  \\  \hline
$Y_{SU(2N-2)}^{bare}$ & \tiny \begin{tabular}{c}$U(1)_2$ charge: \\ $-2(N-1)(2N-6)$　\end{tabular}&1&1&$4-4N$&$-4$&2  \\
 \small  $Y^{d}:=Y_{SU(2N-2)}^{bare} A^{2N-6}$ &1&\scriptsize $(2N-6)$-th symm.&1&$-2N-2$&$-4$&2  \\
 \hline
  \end{tabular}}
  \end{center}\label{SU(2N)2anti04}
\end{table}

\subsection{$SU(2N)$ with $\protect\Young[-0.5]{11}+2 \, \protect\Young[0]{1}+2N \, \overline{\protect\Young[0]{1}}$}
Next, we will study the s-confinement phases with a single anti-symmetric matter. The first example is a 3d $\mathcal{N}=2$ $SU(2N)$ gauge theory with an anti-symmetric tensor, two fundamental matters and $2N$ anti-fundamental matters. This case was studied in \cite{Nii:2016jzi} by using the de-confinement method. The elementary matter fields and their quantum numbers are summarized in Table \ref{SU(2N)122N}. Since the theory is ``chiral,'' the bare Coulomb branch $Y^{bare}_{SU(2N-2)}$ must be dressed. In this example, we can define two types of dressed Coulomb branch operators:
\begin{align}
Y^{dressed}_{A}&:= Y^{bare}_{SU(2N-2)}  \left(  {\tiny \yng(1,1)}_{\,0,-2} \right)^{N-1} \sim Y^{bare}_{SU(2N-2)} A^{N-1}\\
Y^{dressed}_{A Q}&:=   Y^{bare}_{SU(2N-2)}  \left(  {\tiny \yng(1,1)}_{\,0,-2} \right)^{N-2} \left(  {\tiny \yng(1)}_{\, 0,-1} \right)^2 \sim Y^{bare}_{SU(2N-2)} A^{N-2} Q^2,
\end{align}
where the color indices are contracted by the epsilon tensor of the $SU(2N-2)$ gauge group.
The confinement phase is described by the gauge invariant operators listed in  Table \ref{SU(2N)122N} and the superpotential
\begin{align}
W=Y^{dressed}_{A} \left[ \bar{B}_1^{N-1} M^2 +B_{N-1} \bar{B}  \right] + Y^{dressed}_{A Q} \left[ \bar{B}_1^N +T \bar{B} \right].
\end{align}

We can connect this low-energy description to a known s-confinement phase. Let us, for example,  introduce a non-zero vev to $T$. The theory is higgsed to a 3d $\mathcal{N}=2$ $USp(2N)$ gauge theory with $2N+2$ fundamental matters, which exhibits s-confinement \cite{Karch:1997ux, Aharony:1997gp}. On the dual (confining) side, the two fields $Y^{dressed}_{A Q}$ and $\bar{B}$ become massive due to the vev of $T$. The resulting superpotential can be brought together into the $USp(2N)$ confining superpotential \cite{Karch:1997ux, Aharony:1997gp}. Alternatively, by introducing a rank-one vev to $M$, the theory flows to a 3d $\mathcal{N}=2$ $SU(2N-1)$ gauge theory with an anti-symmetric tensor, two fundamental matters and $2N-1$ anti-fundamental matters, which will be discussed in Section 3 and exhibit s-confinement.

\begin{table}[H]\caption{3d $\mathcal{N}=2$ $SU(2N)$ with ${\tiny \protect\yng(1,1)} +2\, {\tiny \protect\yng(1)}+2N \,{\tiny \overline{\protect\yng(1)}}$} 
\begin{center}
\scalebox{0.84}{
  \begin{tabular}{|c||c||c|c|c|c|c|c| } \hline
  &$SU(2N)$&$SU(2)$&$SU(2N)$&$U(1)$&$U(1)$&$U(1)$&$U(1)_R$  \\ \hline
 $A$ &${\tiny \yng(1,1)}$&1&1&1&0&0&$0$ \\
 $Q$ & ${\tiny \yng(1)}$ & ${\tiny \yng(1)}$&1&0&1&0&$0$ \\
$\tilde{Q}$  &${\tiny \overline{\yng(1)}}$&1&${\tiny \yng(1)}$&0&$0$&1&$0$ \\  \hline
$M:=Q\tilde{Q}$&1&${\tiny \yng(1)}$&${\tiny \yng(1)}$&0&1&1&$0$ \\ 
$T:=A^N$&1&1&1&$N$&0&0&$0$ \\
$B_{N-1}:=A^{N-1}Q^2$&1&1&1&$N-1$&2&0&$0$  \\
$\bar{B}:= \tilde{Q}^{2N}$&1&1&1&$0$&0&$2N$&$0$  \\
$\bar{B}_1:=A \tilde{Q}^{2}$&1&1&${\tiny \yng(1,1)}$&1&0&2&$0$  \\  \hline
$Y^{bare}$&\scriptsize $U(1)_2$ charge: $2(N-1)$&1&1&$-2(N-1)$&$-2$&$-2N$&$2$  \\
$Y^{dressed}_{A}:=Y^{bare}_{SU(2N-2)} A^{N-1}$&1&1&1&$-N+1$&$-2$&$-2N$&$2$  \\
$Y^{dressed}_{A Q}:=Y^{bare}_{SU(2N-2)} A^{N-2} Q^2$&1&1&1&$-N$&0&$-2N$&$2$  \\
 \hline
  \end{tabular}}
  \end{center}\label{SU(2N)122N}
\end{table}

\subsection{$SU(2N)$ with $\protect\Young[-0.5]{11}+3 \, \protect\Young[0]{1}+(2N-1) \, \overline{\protect\Young[0]{1}}$}
The UV description is a 3d $\mathcal{N}=2$ $SU(2N)$ gauge theory with an anti-symmetric tensor, three fundamental matters and $2N-1$ anti-fundamental matters. The elementary fields and their quantum numbers are summarized in Table \ref{SU(2N)132Nm1}. This theory is ``vector-like'' in a sense that the corresponding 4d theory has no chiral anomaly for the gauge symmetry. Therefore, the bare Coulomb branch operator $Y^{bare}_{SU(2N-2)}$ is gauge-invariant. The low-energy dynamics is described by the moduli operators defined in Table \ref{SU(2N)132Nm1} and the superpotential
\begin{align}
W= Y^{bare}_{SU(2N-2)}  \left[MB_{N-1}\bar{B}_1^{N-1} + T M^3 \bar{B}_1^{N-2}   \right].
\end{align}

From this theory, we can derive a similar s-confinement phase for an $SU(2N-1)$ gauge group as follows. By introducing a rank-one vev to $M$, the theory flows to a 3d $\mathcal{N}=2$ $SU(2N-1)$ gauge theory with an anti-symmetric tensor, three fundamental matters and $2N-2$ anti-fundamental matters, which will be discussed in Section 3 and exhibit s-confinement.

\begin{table}[H]\caption{3d $\mathcal{N}=2$ $SU(2N)$ with ${\tiny \protect\yng(1,1)} +3\, {\tiny \protect\yng(1)}+ (2N-1) \,{\tiny \overline{\protect\yng(1)}}$} 
\begin{center}
\scalebox{0.93}{
  \begin{tabular}{|c||c||c|c|c|c|c|c| } \hline
  &$SU(2N)$&$SU(3)$&$SU(2N-1)$&$U(1)$&$U(1)$&$U(1)$&$U(1)_R$  \\ \hline
 $A$ &${\tiny \yng(1,1)}$&1&1&1&0&0&$0$ \\
 $Q$ & ${\tiny \yng(1)}$ & ${\tiny \yng(1)}$&1&0&1&0&$0$ \\
$\tilde{Q}$  &${\tiny \overline{\yng(1)}}$&1&${\tiny \yng(1)}$&0&$0$&1&$0$ \\  \hline
$M:=Q\tilde{Q}$&1&${\tiny \yng(1)}$&${\tiny \yng(1)}$&0&1&1&$0$ \\ 
$T:=A^N$&1&1&1&$N$&0&0&$0$ \\
$B_{N-1}:=A^{N-1}Q^2$&1&${\tiny \overline{\yng(1)}}$&1&$N-1$&2&0&$0$  \\
$\bar{B}_1:=A \tilde{Q}^{2}$&1&1&${\tiny \yng(1,1)}$&1&0&2&$0$  \\  \hline
$Y^{bare}_{SU(2N-2)}$&1&1&1&$-2(N-1)$&$-3$&$-2N+1$&$2$  \\
 \hline
  \end{tabular}}
  \end{center}\label{SU(2N)132Nm1}
\end{table}

\subsection{$SU(2N)$ with $\protect\Young[-0.5]{11}+4 \, \protect\Young[0]{1}+(2N-2) \, \overline{\protect\Young[0]{1}}$}
Let us consider the 3d $\mathcal{N}=2$ $SU(2N)$ gauge theory with an anti-symmetric matter, four fundamental matters and $2N-2$ anti-fundamental matters. The bare Coulomb branch $Y^{bare}_{SU(2N-2)}$ has a non-zero $U(1)_2$ charge $-2(N-1)$ and this can be canceled as
\begin{align}
Y^{dressed}_{A}  &:=Y^{bare}_{SU(2N-2)} 1_{0,2(N-1)}  \sim  Y^{bare}_{SU(2N-2)} A  \\
Y^{dressed}_{\tilde{Q}}  &:=Y^{bare}_{SU(2N-2)} \left( {\tiny \overline{ \yng(1)}}_{\, 0,1}  \right)^{2N-2}  \sim Y^{bare}_{SU(2N-2)} \tilde{Q}^{2N-2},
\end{align}
where the color and flavor indices of $\tilde{Q}^{2N-2}$ are totally anti-symmetrized and the dressed operators have no flavor index. The low-energy effective description is given by a non-gauge theory with the gauge singlets defined in Table \ref{SU(2N)122Nm2}. The confining superpotential becomes
\begin{align}
W= Y_A^{dressed} \left[ \bar{B}_1^{N-2} M^2 B_{N-1} +T \bar{B}_1^{N-3} M^4 \right]  +Y_{\tilde{Q}}^{dressed} \left[ B_{N-1}^2 +TB_{N-2} \right].
\end{align}
As a consistency check, we can flow to a similar s-confinement with an $SU(2N-1)$ gauge group. By introducing a rank-one vev to $M$, the theory flows to a 3d $\mathcal{N}=2$ $SU(2N-1)$ gauge theory with an anti-symmetric tensor, four fundamental matters and $2N-3$ anti-fundamental matters, which will be discussed in Section 3 and exhibit s-confinement.

\begin{table}[H]\caption{3d $\mathcal{N}=2$ $SU(2N)$ with ${\tiny \protect\yng(1,1)} +4\, {\tiny \protect\yng(1)}+ (2N-2) \,{\tiny \overline{\protect\yng(1)}}$} 
\begin{center}
\scalebox{0.77}{
  \begin{tabular}{|c||c||c|c|c|c|c|c| } \hline
  &$SU(2N)$&$SU(4)$&$SU(2N-2)$&$U(1)$&$U(1)$&$U(1)$&$U(1)_R$  \\ \hline
 $A$ &${\tiny \yng(1,1)}$&1&1&1&0&0&$0$ \\
 $Q$ & ${\tiny \yng(1)}$ & ${\tiny \yng(1)}$&1&0&1&0&$0$ \\
$\tilde{Q}$  &${\tiny \overline{\yng(1)}}$&1&${\tiny \yng(1)}$&0&$0$&1&$0$ \\  \hline
$M:=Q\tilde{Q}$&1&${\tiny \yng(1)}$&${\tiny \yng(1)}$&0&1&1&$0$ \\ 
$T:=A^N$&1&1&1&$N$&0&0&$0$ \\
$B_{N-1}:=A^{N-1}Q^2$&1&${\tiny \yng(1,1)}$&1&$N-1$&2&0&$0$  \\
$B_{N-2}:=A^{N-2}Q^4$&1&1&1&$N-2$&4&0&$0$  \\
$\bar{B}_1:=A \tilde{Q}^{2}$&1&1&${\tiny \yng(1,1)}$&1&0&2&$0$  \\  \hline
$Y^{bare}_{SU(2N-2)}$&\scriptsize $U(1)_2$ charge: $-2(N-1)$&1&1&$-2(N-2)$&$-4$&$-(2N-2)$&$2$  \\
$Y^{dressed}_{A}:=Y^{bare}_{SU(2N-2)} A$&1&1&1&$-2N+3$&$-4$&$-(2N-2)$&$2$  \\
$Y^{dressed}_{\tilde{Q}}:=Y^{bare}_{SU(2N-2)} \tilde{Q}^{2N-2} $&1&1&1&$-(2N-2)$&$-4$&$0$&$2$  \\
 \hline
  \end{tabular}}
  \end{center}\label{SU(2N)122Nm2}
\end{table}

\subsection{$SU(2N)$ with $\protect\Young[-0.5]{11}+\overline{\protect\Young[-0.5]{11}}+2 ( \protect\Young[0]{1}+ \overline{\protect\Young[0]{1}} )$}
We next consider the 3d $\mathcal{N}=2$ $SU(2N)$ gauge theory with an anti-symmetric flavor and two (anti-)fundamental flavors. This theory was studied in \cite{Nii:2016jzi, Csaki:2014cwa}. The theory is ``vector-like'' in a sense that the corresponding 4d theory has no gauge anomaly. Therefore, the bare Coulomb branch operator $Y_{SU(2N-2)}^{bare}$ is gauge-invariant. 
Along the Coulomb branch spanned by $Y_{SU(2N-2)}^{bare}$, the second-order anti-symmetric representations reduce to two different massless components
\begin{align}
{\tiny \yng(1,1)} \ni {\tiny \yng(1,1)}_{\,0,-2}+1_{0,2(N-1)},~~~{\tiny \overline{\yng(1,1)}} \ni {\tiny \overline{\yng(1,1)}}_{\,0,2} +1_{0,-2(N-1)}.
\end{align}
As a result, we can define the following dressed operators
\begin{align}
Y_{a=0,\cdots,N-1}&:=Y_{SU(2N-2)}^{bare}  \left(1_{0,2(N-1)} 1_{0,-2(N-1)} \right)^a   \\
&\sim Y_{SU(2N-2)}^{bare} (A \tilde{A})^a,~~~~a=0,\cdots,N-1.
\end{align}
These should be recognized as the moduli coordinates which are independent of $Y_{SU(2N-2)}^{bare}  T_a$. The low-energy dynamics is described by the gauge invariant operators listed in Table \ref{SU(2N)antiflavor22}. We will not explicitly write down the confining superpotential but, for each $N$, one can write down it. By introducing a rank-one vev to $M_0$, the theory flows to a 3d $\mathcal{N}=2$ $SU(2N-1)$ gauge theory with an anti-symmetric flavor and two (anti-)fundamental flavors, which will be discussed in Section 3 and exhibit s-confinement.

\begin{table}[H]\caption{3d $\mathcal{N}=2$ $SU(2N)$ with $\protect\Young[-0.5]{11}+\overline{\protect\Young[-0.5]{11}}+2 ( \protect\Young[0]{1}+ \overline{\protect\Young[0]{1}} )$} 
\begin{center}
\scalebox{0.8}{
  \begin{tabular}{|c||c||c|c|c|c|c|c|c| } \hline
  &$SU(2N)$&$SU(2)$&$SU(2)$&$U(1)$&$U(1)$&$U(1)$&$U(1)$&$U(1)_R$  \\ \hline
 $A$ &${\tiny \yng(1,1)}$&1&1&1&0&0&0&$0$ \\
  $\tilde{A}$ &${\tiny \overline{\yng(1,1)}}$&1&1&0&1&0&0&$0$ \\
 $Q$ & ${\tiny \yng(1)}$ & ${\tiny \yng(1)}$&1&0&0&1&0&$0$ \\
$\tilde{Q}$  &${\tiny \overline{\yng(1)}}$&1&${\tiny \yng(1)}$&0&$0$&0&1&$0$ \\  \hline
$M_{k=0,\cdots,N-1}:=Q(A\tilde{A})^k \tilde{Q}$&1&${\tiny \yng(1)}$&${\tiny \yng(1)}$&$k$&$k$&1&1&0  \\
$H_{m=0,\cdots,N-2}:=\tilde{A}(A \tilde{A})^m Q^2$&1&1&1&$m$&$m+1$&2&0&0  \\
$\bar{H}_{m=0,\cdots,N-2}:=A(A \tilde{A})^m \tilde{Q}^2 $&1&1&1&$m+1$&$m$&0&2&0  \\
$B_N:=A^N$&1&1&1&$N$&0&0&0&0  \\
$\bar{B}_{N}:=\tilde{A}^N$&1&1&1&0&$N$&0&0&0  \\
$B_{N-1}:=A^{N-1}Q^2$&1&1&1&$N-1$&0&2&0&0  \\
$\bar{B}_{N-1}:=\tilde{A}^{N-1} \tilde{Q}^2$&1&1&1&0&$N-1$&0&2&0  \\
$T_{n=1,\cdots,N-1}:=(A \tilde{A})^n$&1&1&1&$n$&$n$&0&0&0  \\  \hline
$Y_{a=0,\cdots,N-1}:=Y_{SU(2N-2)}^{bare} (A \tilde{A})^a$&1&1&1&$2-2N+a$&$2-2N+a$&$-2$&$-2$&2  \\
 \hline
  \end{tabular}}
  \end{center}\label{SU(2N)antiflavor22}
\end{table}

\subsection{$SU(2N)$ with $\protect\Young[-0.5]{11}+\overline{\protect\Young[-0.5]{11}}+3\,  \protect\Young[0]{1}+ \overline{\protect\Young[0]{1}}$}
The UV description is a 3d $\mathcal{N}=2$ $SU(2N)$ gauge theory with an anti-symmetric flavor and three fundamental matters and an anti-fundamental matter. Table \ref{SU(2N)antiflavor31} summarizes the quantum numbers of the elementary fields and the moduli operators. The $U(1)_2$ charge of the bare Coulomb branch $Y_{SU(2N-2)}^{bare}$ is $-2(N-1)$ and the dressed operators are defined by
\begin{align}
Y^{dressed}_{a} &:=Y_{SU(2N-2)}^{bare} 1_{0,2(N-1)}  (1_{0,2(N-1)}1_{0,-2(N-1)})^a \nonumber \\
&\quad \sim Y_{SU(2N-2)}^{bare} A(A \tilde{A})^a,~~~a=0,\cdots,N-2 \\
Y_{\tilde{A}}^{dressed} &:=Y_{SU(2N-2)}^{bare}  \left(  {\tiny \overline{\yng(1,1)}}_{\,0,2} \right)^{N-1}    \sim Y_{SU(2N-2)}^{bare}\tilde{A}^{N-1}.
\end{align}
Notice that a dressed operator such as
\begin{align}
 Y_{SU(2N-2)}^{bare}  \left(  {\tiny \overline{\yng(1,1)}}_{\,0,2} \right)^{N-1} (1_{0,2(N-1)}1_{0,-2(N-1)})
\end{align}
is identified with $Y^{dressed}_{a=0} \bar{B}_N$ and cannot be an independent operator. The low-energy dynamics is described by the gauge-invariant moduli fields in Table \ref{SU(2N)antiflavor31}. By introducing a rank-one vev to $M_0$, the theory flows to a 3d $\mathcal{N}=2$ $SU(2N-1)$ gauge theory with an anti-symmetric flavor, three fundamental matters and an anti-fundamental matter, which will be discussed in Section 3 and again exhibit s-confinement.

\begin{table}[H]\caption{3d $\mathcal{N}=2$ $SU(2N)$ with $\protect\Young[-0.5]{11}+\overline{\protect\Young[-0.5]{11}}+3\,  \protect\Young[0]{1}+ \overline{\protect\Young[0]{1}}$} 
\begin{center}
\scalebox{0.79}{
  \begin{tabular}{|c||c||c|c|c|c|c|c| } \hline
  &$SU(2N)$&$SU(3)$&$U(1)$&$U(1)$&$U(1)$&$U(1)$&$U(1)_R$  \\ \hline
 $A$ &${\tiny \yng(1,1)}$&1&1&0&0&0&$0$ \\
  $\tilde{A}$ &${\tiny \overline{\yng(1,1)}}$&1&0&1&0&0&$0$ \\
 $Q$ & ${\tiny \yng(1)}$ & ${\tiny \yng(1)}$&0&0&1&0&$0$ \\
$\tilde{Q}$  &${\tiny \overline{\yng(1)}}$&1&0&$0$&0&1&$0$ \\  \hline
$M_{k=0,\cdots,N-1}:=Q(A\tilde{A})^k \tilde{Q}$&1&${\tiny \yng(1)}$&$k$&$k$&1&1&0  \\
$H_{m=0,\cdots,N-2}:=\tilde{A}(A \tilde{A})^m Q^2$&1&${\tiny \overline{\yng(1)}}$&$m$&$m+1$&2&0&0  \\
$B_N:=A^N$&1&1&$N$&0&0&0&0  \\
$\bar{B}_{N}:=\tilde{A}^N$&1&1&0&$N$&0&0&0  \\
$B_{N-1}:=A^{N-1}Q^2$&1&${\tiny \overline{\yng(1)}}$&$N-1$&0&2&0&0  \\
$T_{n=1,\cdots,N-1}:=(A \tilde{A})^n$&1&1&$n$&$n$&0&0&0  \\  \hline
$Y_{SU(2N-2)}^{bare}$&\scriptsize $U(1)_2$ charge:$-2(N-1)$&1&$2-2N$&$2-2N$&$-3$&$-1$&2  \\
$Y^{dressed}_{a=0,\cdots,N-2}:=Y_{SU(2N-2)}^{bare} A(A \tilde{A})^a$&1&1&$3-2N+a$&$2-2N+a$&$-3$&$-1$&2  \\
$Y_{\tilde{A}}^{dressed}:=Y_{SU(2N-2)}^{bare}\tilde{A}^{N-1}$&1&1&$2-2N$&$1-N$&$-3$&$-1$&2  \\
 \hline
  \end{tabular}}
  \end{center}\label{SU(2N)antiflavor31}
\end{table}

\subsection{$SU(2N)$ with $\protect\Young[-0.5]{11}+\overline{\protect\Young[-0.5]{11}}+4\,  \protect\Young[0]{1}$}
The final example is a 3d $\mathcal{N}=2$ $SU(2N)$ gauge theory with an anti-symmetric flavor and four fundamental matters. Table \ref{SU(2N)antiflavor40} summarizes the quantum numbers of the elementary fields and the moduli coordinates. Since the matter content of the (anti-)fundamental representations is ``chiral,'' the bare Coulomb branch operator $Y_{SU(2N-2)}^{bare}$ has a non-zero $U(1)_2$ charge. The dressed (gauge-invariant) operators are defined by
\begin{align}
Y^{dressed}_{a=0,\cdots,N-2} &:=Y_{SU(2N-2)}^{bare} (1_{0,2(N-1)})^2 (1_{0,2(N-1)}1_{0,-2(N-1)})^a \sim Y_{SU(2N-2)}^{bare} A^2(A \tilde{A})^a \\
Y_{\tilde{A}}^{dressed}& := Y_{SU(2N-2)}^{bare}  \left(  {\tiny \overline{\yng(1,1)}}_{\,0,2} \right)^{2(N-1)}   \sim Y_{SU(2N-2)}^{bare} (\tilde{A}^{N-1})^2 \\
Y_{A\tilde{A}}^{dressed} &:=  Y_{SU(2N-2)}^{bare}1_{0,2(N-1)}  \left(  {\tiny \overline{\yng(1,1)}}_{\,0,2} \right)^{N-1}   \sim Y_{SU(2N-2)}^{bare} (\tilde{A}^{N-1})A.
\end{align}
Notice that the dressed operator such as
\begin{align}
 Y_{SU(2N-2)}^{bare}  \left(  {\tiny \overline{\yng(1,1)}}_{\,0,2} \right)^{2(N-1)} (1_{0,2(N-1)}1_{0,-2(N-1)})
\end{align}
is identified with $Y^{dressed}_{A \tilde{A}} \bar{B}_N$ and cannot be an independent operator. The low-energy dynamics is described by the gauge-invariant fields in Table \ref{SU(2N)antiflavor40}

\begin{table}[H]\caption{3d $\mathcal{N}=2$ $SU(2N)$ with $\protect\Young[-0.5]{11}+\overline{\protect\Young[-0.5]{11}}+4\,  \protect\Young[0]{1}$} 
\begin{center}
\scalebox{0.8}{
  \begin{tabular}{|c||c||c|c|c|c|c| } \hline
  &$SU(2N)$&$SU(4)$&$U(1)$&$U(1)$&$U(1)$&$U(1)_R$  \\ \hline
 $A$ &${\tiny \yng(1,1)}$&1&1&0&0&$0$ \\
  $\tilde{A}$ &${\tiny \overline{\yng(1,1)}}$&1&0&1&0&$0$ \\
 $Q$ & ${\tiny \yng(1)}$ & ${\tiny \yng(1)}$&0&0&1&$0$ \\ \hline
$H_{m=0,\cdots,N-2}:=\tilde{A}(A \tilde{A})^m Q^2$&1&${\tiny \yng(1,1)}$&$m$&$m+1$&2&0  \\
$B_N:=A^N$&1&1&$N$&0&0&0  \\
$B_{N-1}:=A^{N-1}Q^2$&1&${\tiny \yng(1,1)}$&$N-1$&0&2&0  \\
$B_{N-2}:=A^{N-2}Q^4$&1&$1$&$N-2$&0&4&0  \\
$\bar{B}_{N}:=\tilde{A}^N$&1&1&0&$N$&0&0  \\
$T_{n=1,\cdots,N-1}:=(A \tilde{A})^n$&1&1&$n$&$n$&0&0  \\  \hline
$Y_{SU(2N-2)}^{bare}$&\scriptsize $U(1)_2$ charge:$-4(N-1)$&1&$2-2N$&$2-2N$&$-4$&2  \\
$Y^{dressed}_{a=0,\cdots,N-2}:=Y_{SU(2N-2)}^{bare} A^2(A \tilde{A})^a$&1&1&$4-2N+a$&$2-2N+a$&$-4$&2  \\
$Y_{\tilde{A}}^{dressed}:=Y_{SU(2N-2)}^{bare} (\tilde{A}^{N-1})^2$&1&1&$2-2N$&0&$-4$&2  \\
$Y_{A\tilde{A}}^{dressed}:=Y_{SU(2N-2)}^{bare} (\tilde{A}^{N-1})A$&1&1&$3-2N$&$1-N$&$-4$&2  \\
 \hline
  \end{tabular}}
  \end{center}\label{SU(2N)antiflavor40}
\end{table}

\section{$SU(2N+1)$ gauge theories}
In this section, we will study the s-confinement phases in the 3d $\mathcal{N}=2$ $SU(2N+1)$ gauge theories with anti-symmetric and (anti-)fundamental matters. The analysis of the Coulomb branch is very similar to the previous one with a small modification. When the bare Coulomb branch operator denoted by $Y^{bare}_{SU(2N-1)}$ obtains an expectation value, the gauge group is spontaneously broken to
\begin{align}
SU(2N+1) & \rightarrow SU(2N-1) \times U(1)_1 \times U(1)_2 \\
{\tiny \yng(1)} &  \rightarrow {\tiny \yng(1)}_{\, 0,-2} + \mathbf{1}_{1,2N-1}+ \mathbf{1}_{-1,2N-1} \\
{\tiny \overline{\yng(1)}} &  \rightarrow {\tiny \overline{\yng(1)}}_{\, 0,2} + \mathbf{1}_{-1,-(2N-1)}+ \mathbf{1}_{1,-(2N-1)} \\
{\tiny \yng(1,1)} &  \rightarrow {\tiny \yng(1,1)}_{\, 0,-4}+{\tiny \yng(1)}_{\, 1,2N-3}+{\tiny \yng(1)}_{-1,2N-3} +\mathbf{1}_{0,4N-2} \\
{\tiny \overline{\yng(1,1)}} &  \rightarrow {\tiny \overline{\yng(1,1)}}_{\, 0,4}+{\tiny \overline{\yng(1)}}_{\, -1,-(2N-3)}+{\tiny \overline{\yng(1)}}_{1,-(2N-3)} +\mathbf{1}_{0,-(4N-2)}.
\end{align}
When the theory includes $F$ fundamental matters, $\bar{F}$ anti-fundamental matters, $F_A$ anti-symmetric tensors and $\bar{F}_A$ anti-symmetric-bar tensors, the $U(1)_2$ charge of $Y^{bare}_{SU(2N-1)}$ becomes 
\begin{align}
U(1)_2[Y^{bare}_{SU(2N-1)}]=-(2N-1)(F  -\bar{F}) -(2N-1) (2N-3) (F_A -\bar{F}_A)
\end{align}
For the ``chiral'' matter contents where the corresponding 4d theory has a chiral gauge anomaly, the bare Coulomb branch operator must be dressed by matter fields. In the following subsections, we will list various examples of the s-confinement phases.

\subsection{$SU(2N+1)$ with $2\, \protect\Young[-0.5]{11}+4 \, \protect\Young[0]{1}$}
The first example is a 3d $\mathcal{N}=2$ $SU(2N+1)$ gauge theory with two anti-symmetric matters and four fundamental matters. Table \ref{SU(2N+1)2anti40} summarizes the quantum numbers of the elementary fields and the moduli coordinates. Since the theory is ``chiral,'' the bare Coulomb branch $ Y_{SU(2N-1)}^{bare} $ must be dressed by the massless matter components
\begin{align}
Y^{d}:=  Y_{SU(2N-1)}^{bare}  \left( \mathbf{1}_{0,4N-2} \right)^{2N-1}   \sim Y_{SU(2N-1)}^{bare} A^{2N-1},
\end{align}
where the flavor indices of $A^{2N-1}$ are symmetrized. 
The low-energy dynamics is described by the gauge invariant operators in Table \ref{SU(2N+1)2anti40} and a confining superpotential
\begin{align}
W=Y^d  T_N T_{N-1}.
\end{align}
For $N=1$, the theory reduces to the 3d $\mathcal{N}=2$ $SU(3)$ gauge theory with four fundamental and two anti-fundamental matters, which was studied in \cite{Nii:2018bgf} and exhibits an s-confinement phase. The case with $N=2$ will be individually discussed in Section 5.

\begin{table}[H]\caption{3d $\mathcal{N}=2$ $SU(2N+1)$ with $2\, {\tiny \protect\yng(1,1)} +4\, {\tiny \protect\yng(1)}$} 
\begin{center}
\scalebox{1}{
  \begin{tabular}{|c||c||c|c|c|c|c| } \hline
  &$SU(2N+1)$&$SU(2)$&$SU(4)$&$U(1)$&$U(1)$&$U(1)_R$  \\ \hline
 $A$ &${\tiny \yng(1,1)}$& ${\tiny \yng(1)}$ &1&1&0&0 \\
 $Q$ & ${\tiny \yng(1)}$ &1& ${\tiny \yng(1)}$&0&1&0 \\  \hline
 $T_N:=A^N Q$&1&\scriptsize $N$-th symm.&${\tiny \yng(1)}$&$N$&1&0  \\
 $T_{N-1}:=A^{N-1}Q^3$&1&\scriptsize $(N-1)$-th symm.&${\tiny \overline{\yng(1)}}$&$N-1$&3&0  \\ \hline
$Y_{SU(2N-1)}^{bare}$ &\tiny \begin{tabular}{c}$U(1)_2$ charge: \\ $-2(2N-1)^2$　\end{tabular}&1&1&$2-4N$&$-4$&2  \\
 \small  $Y^{d}:=Y_{SU(2N-1)}^{bare} A^{2N-1}$ &1&\scriptsize $(2N-1)$-th symm.&1&$1-2N$&$-4$&2  \\
 \hline
  \end{tabular}}
  \end{center}\label{SU(2N+1)2anti40}
\end{table}

\subsection{$SU(2N+1)$ with $2\, \protect\Young[-0.5]{11}+3 \, \protect\Young[0]{1}+  \overline{\protect\Young[0]{1}}$}
The second example is a 3d $\mathcal{N}=2$ $SU(2N+1)$ gauge theory with two anti-symmetric matters, three fundamental matters and a single anti-fundamental matter. The elementary fields and their quantum numbers are summarized in Table \ref{SU(2N+1)2anti31}. The bare Coulomb branch $Y_{SU(2N-1)}^{bare}$ has a non-zero $U(1)_2$ charge $-2(2N-1)(2N-2)$ and the dressed operator is defined by
\begin{align}
Y^d:= Y_{SU(2N-1)}^{bare}  \left( \mathbf{1}_{0,4N-2} \right)^{2N-2} \sim Y_{SU(2N-1)}^{bare} A^{2N-2}, 
\end{align}
where the flavor indices of $A^{2N-2}$ are totally symmetrized. The low-energy effective theory is described by the gauge invariant chiral superfields defined in Table \ref{SU(2N+1)2anti31} and a confining superpotential
\begin{align}
W=Y^d \left(  MT_N^2+ T_{N-1} P_1 +T_NP_2 \right).
\end{align}

As a simple consistency check, for $N=1$, we don't have to dress the bare Coulomb branch and the theory becomes a 3d $\mathcal{N}=2$ $SU(3)$ gauge theory with three (anti-)fundamental flavors, which shows s-confinement \cite{Aharony:1997bx}. By introducing a rank-one vev to $M$, the theory flows to a 3d $\mathcal{N}=2$ $SU(2N)$ gauge theory with two anti-symmetric tensors and four fundamental matters, which exhibits s-confinement as studied in the previous section.

\begin{table}[H]\caption{3d $\mathcal{N}=2$ $SU(2N+1)$ with $2\, {\tiny \protect\yng(1,1)} +3\, {\tiny \protect\yng(1)}+ {\tiny \overline{\protect\yng(1)}}$} 
\begin{center}
\scalebox{0.88}{
  \begin{tabular}{|c||c||c|c|c|c|c|c| } \hline
  &$SU(2N+1)$&$SU(2)$&$SU(3)$&$U(1)$&$U(1)$&$U(1)$&$U(1)_R$  \\ \hline
 $A$ &${\tiny \yng(1,1)}$& ${\tiny \yng(1)}$ &1&1&0&0&0 \\
 $Q$ & ${\tiny \yng(1)}$ &1& ${\tiny \yng(1)}$&0&1&0&0 \\  
$\tilde{Q}$ &${\tiny \overline{ \yng(1)}}$&1&1&0&0&1&0  \\  \hline
$M:=Q \tilde{Q}$ &1&1&${\tiny \yng(1)}$ &0&1&1&0  \\
$T_N:=A^NQ$ &1&\scriptsize$N$-th symm.& ${\tiny \yng(1)}$ &$N$&1&0&0  \\
 $T_{N-1}:=A^{N-1}Q^3$ &1&\scriptsize$(N-1)$-th symm.&1&$N-1$&3&0&0  \\
 $P_1:=A^{N}(A\tilde{Q})$  &1&\scriptsize$(N-1)$-th symm.&1&$N+1$&0&1&0  \\
$P_2:=A^{N-1}(A \tilde{Q})Q^2$ &1&\scriptsize$(N-2)$-th symm.&${\tiny \overline{\yng(1)}}$&$N$&2&1&0  \\  \hline
  $Y_{SU(2N-1)}^{bare}$&\tiny \begin{tabular}{c}$U(1)_2$ charge: \\ $-2(2N-1)(2N-2)$　\end{tabular}&1&1&$2-4N$&$-3$&$-1$&2 \\
   $Y^d:=Y_{SU(2N-1)}^{bare} A^{2N-2} $&1&\scriptsize$(2N-2)$-th symm.&1&$-2N$&$-3$&$-1$&2  \\  \hline
  \end{tabular}}
  \end{center}\label{SU(2N+1)2anti31}
\end{table}

\subsection{$SU(2N+1)$ with $2\, \protect\Young[-0.5]{11}+2 \, \protect\Young[0]{1}+2 \, \overline{\protect\Young[0]{1}}$}
The third example is a 3d $\mathcal{N}=2$ $SU(2N+1)$ gauge theory with two anti-symmetric matters, two (anti-)fundamental flavors. Table \ref{SU(2N+1)2anti22} summarizes the elementary fields and their quantum numbers. The $U(1)_2$ charge of the bare Coulomb branch operator $Y_{SU(2N-1)}^{bare}$ is $-2(2N-1)(2N-3)$ and the dressed operator is defined by
\begin{align}
Y^d:=Y_{SU(2N-1)}^{bare}  \left( \mathbf{1}_{0,4N-2} \right)^{2N-3}  \sim Y_{SU(2N-1)}^{bare} A^{2N-3},
\end{align}
where the flavor indices of $A^{2N-3}$ are symmetrized. For $N=1$, $Y_{SU(2N-1)}^{bare}$ is positively charged and must be dressed by a massless component of $Q$. The low-energy dynamics is described by the gauge invariant operators in Table \ref{SU(2N+1)2anti22}.
The confining superpotential becomes 
\begin{align}
W=Y^{d} \left(MT_NP_N + \bar{B}T_N^2+T_NT_{N-1}+ P_N P_{N-1} \right).
\end{align}

As a consistency check, for $N=1$, the theory becomes a 3d $\mathcal{N}=2$ $SU(3)$ gauge theory with two fundamental matters and four anti-fundamental matters, which shows s-confinement \cite{Nii:2018bgf}. By introducing a rank-one vev to $M$, the theory flows to a 3d $\mathcal{N}=2$ $SU(2N)$ gauge theory with two anti-symmetric tensors, three fundamental matters and an anti-fundamental matter, which exhibits s-confinement as studied in the previous section.

\begin{table}[H]\caption{3d $\mathcal{N}=2$ $SU(2N+1)$ with $2\, {\tiny \protect\yng(1,1)} +2\, {\tiny \protect\yng(1)}+2 \, {\tiny \overline{\protect\yng(1)}}$} 
\begin{center}
\scalebox{0.81}{
  \begin{tabular}{|c||c||c|c|c|c|c|c|c| } \hline
  &$SU(2N+1)$&$SU(2)$&$SU(2)$&$SU(2)$&$U(1)$&$U(1)$&$U(1)$&$U(1)_R$  \\ \hline
 $A$ &${\tiny \yng(1,1)}$& ${\tiny \yng(1)}$ &1&1&1&0&0&0 \\
 $Q$ & ${\tiny \yng(1)}$ &1& ${\tiny \yng(1)}$&1&0&1&0&0 \\  
$\tilde{Q}$ &${\tiny \overline{ \yng(1)}}$&1&1&${\tiny \yng(1)}$&0&0&1&0  \\  \hline
$M:=Q \tilde{Q}$ &1&1&${\tiny \yng(1)}$&${\tiny \yng(1)}$ &0&1&1&0  \\
$\bar{B}:=A \tilde{Q}^2$ &1& ${\tiny \yng(1)}$&1&1&1&0&2&0  \\
$T_N:=A^NQ$ &1&\scriptsize$N$-th symm.& ${\tiny \yng(1)}$ &1&$N$&1&0&0  \\
 $T_{N-1}:=A^{N-1}(A\tilde{Q})^2Q$ &1&\scriptsize$(N-3)$-th symm.& ${\tiny \yng(1)}$&1&$N+1$&1&2&0  \\
 $P_{N}:=A^{N}(A \tilde{Q})$&1&\scriptsize$(N-1)$-th symm.&1&${\tiny \yng(1)}$&$N+1$&0&1&0 \\
$P_{N-1}:=A^{N-1} (A\tilde{Q})Q^2$ &1&\scriptsize$(N-2)$-th symm.&1&${\tiny \yng(1)}$&$N$&2&1&0 \\   \hline
 $Y_{SU(2N-1)}^{bare}$ &\tiny \begin{tabular}{c}$U(1)_2$ charge: \\ $-2(2N-1)(2N-3)$　\end{tabular}&1&1&1&$2-4N$&$-2$&$-2$&2 \\ 
\small  $Y^d:=Y_{SU(2N-1)}^{bare} A^{2N-3} $&1&\scriptsize$(2N-3)$-th symm.&1&1&$-1-2N$&$-2$&$-2$&2 \\  \hline
  \end{tabular}}
  \end{center}\label{SU(2N+1)2anti22}
\end{table}

\subsection{$SU(2N+1)$ with $2\, \protect\Young[-0.5]{11}+  \protect\Young[0]{1}+3 \, \overline{\protect\Young[0]{1}}\,$, $(N>1)$}
The fourth example is a 3d $\mathcal{N}=2$ $SU(2N+1)$ gauge theory with two anti-symmetric tensors, one fundamental matter and three anti-fundamental matters. We here assume $N>1$. Since the theory is ``chiral,'' the bare Coulomb brach $Y_{SU(2N-1)}^{bare}$ is not gauge-invariant. The dressed operator is defined by
\begin{align}
Y^d:= Y_{SU(2N-1)}^{bare}  \left( \mathbf{1}_{0,4N-2} \right)^{2N-4}   \sim Y_{SU(2N-1)}^{bare} A^{2N-4},
\end{align}
where the flavor indices of $A^{2N-4}$ are totally symmetrized.
Table \ref{SU(2N+1)2anti13} summarizes the quantum numbers of the elementary fields and the moduli coordinates. The symmetry argument determines the confining superpotential as
\begin{align}
W=Y^d \left( MP_N^2 +\bar{B}P_N T_N+T_NP_{N-1}+ P_NR \right).
\end{align}
By introducing a rank-one vev to $M$, the theory flows to a 3d $\mathcal{N}=2$ $SU(2N)$ gauge theory with two anti-symmetric tensors and two (anti-)fundamental flavors, which exhibits s-confinement as studied in the previous section.

\begin{table}[H]\caption{3d $\mathcal{N}=2$ $SU(2N+1)$ with $2\, {\tiny \protect\yng(1,1)} + {\tiny \protect\yng(1)}+3\, {\tiny \overline{\protect\yng(1)}}$} 
\begin{center}
\scalebox{0.85}{
  \begin{tabular}{|c||c||c|c|c|c|c|c| } \hline
  &$SU(2N+1)$&$SU(2)$&$SU(3)$&$U(1)$&$U(1)$&$U(1)$&$U(1)_R$  \\ \hline
 $A$ &${\tiny \yng(1,1)}$& ${\tiny \yng(1)}$ &1&1&0&0&0 \\
 $Q$ & ${\tiny \yng(1)}$ &1&1&0&1&0&0 \\  
$\tilde{Q}$ &${\tiny \overline{ \yng(1)}}$&1& ${\tiny \yng(1)}$&0&0&1&0  \\  \hline
$M:=Q \tilde{Q}$ &1&1&${\tiny \yng(1)}$ &0&1&1&0  \\
$\bar{B}:=A \tilde{Q}^2$&1&${\tiny \yng(1)}$&${\tiny \overline{\yng(1)}}$&1&0&2&0  \\
$T_N:=A^N Q$ &1&\scriptsize$N$-th symm.&1&$N$&1&0&0  \\
 $P_N:=A^{N}(A\tilde{Q})$  &1&\scriptsize$(N-1)$-th symm.&${\tiny \yng(1)}$ &$N+1$&0&1&0  \\
  $P_{N-1}:=A^{N-1}(A\tilde{Q})^3$  &1&\scriptsize$(N-4)$-th symm.&1 &$N+2$&0&3&0  \\
$R:=A^{N-1}(A \tilde{Q})^2Q$ &1&\scriptsize$(N-3)$-th symm.&${\tiny \overline{ \yng(1)}}$&$N+1$&1&2&0  \\  \hline
  $Y_{SU(2N-1)}^{bare}$&\tiny \begin{tabular}{c}$U(1)_2$ charge: \\ $-2(2N-1)(2N-4)$　\end{tabular}&1&1&$2-4N$&$-1$&$-3$&2 \\
   $Y^d:=Y_{SU(2N-1)}^{bare} A^{2N-4} $&1&\scriptsize$(2N-4)$-th symm.&1&$-2N-2$&$-1$&$-3$&2  \\  \hline
  \end{tabular}}
  \end{center}\label{SU(2N+1)2anti13}
\end{table}

\subsection{$SU(2N+1)$ with $2\, \protect\Young[-0.5]{11}+ 4\, \overline{\protect\Young[0]{1}}\,$, $(N>1)$}
The fifth example is a 3d $\mathcal{N}=2$ $SU(2N+1)$ gauge theory with two anti-symmetric tensors and four anti-fundamental matters, where $N>1$. The matter fields and their quantum numbers are summarized in Table \ref{SU(2N+1)2anti04}. The $U(1)_2$ charge of the bare Coulomb branch operator $Y_{SU(2N-1)}^{bare}$ is $-2(2N-1)(2N-5)$ and the dressed operator is defined by
\begin{align}
Y^{d}:= Y_{SU(2N-1)}^{bare}\left( \mathbf{1}_{0,4N-2} \right)^{2N-5}   \sim Y_{SU(2N-1)}^{bare} A^{2N-5},
\end{align}
where the flavor indices of $A^{2N-5}$ is totally symmetrized. For $N=2$, $Y_{SU(2N-1)}^{bare}$ has a positive $U(1)_2$ charge which should be canceled by $A^2 \tilde{Q}$, which will be discussed in Section 5. The low-energy dynamics is described by the moduli coordinates listed in Table \ref{SU(2N+1)2anti04}.
The confining superpotential is determined as
\begin{align}
W=Y^d \left( P_N P_{N-1} + \bar{B}T_{N}^2  \right).
\end{align}

\begin{table}[H]\caption{3d $\mathcal{N}=2$ $SU(2N+1)$ with $2\, {\tiny \protect\yng(1,1)} +4\, {\tiny \overline{\protect\yng(1)}}$} 
\begin{center}
\scalebox{0.93}{
  \begin{tabular}{|c||c||c|c|c|c|c| } \hline
  &$SU(2N+1)$&$SU(2)$&$SU(4)$&$U(1)$&$U(1)$&$U(1)_R$  \\ \hline
 $A$ &${\tiny \yng(1,1)}$& ${\tiny \yng(1)}$ &1&1&0&0 \\
 $\tilde{Q}$ & ${\tiny \overline{ \yng(1)}}$ &1& ${\tiny \yng(1)}$&0&1&0 \\  \hline
$\bar{B}:=A\tilde{Q}^2 $&1&${\tiny \yng(1)}$&${\tiny \yng(1,1)}$&1&2&0 \\
$P_N:=A^N(A \tilde{Q})$&1&\scriptsize $N-1$-th symm.&${\tiny \yng(1)}$&$N+1$&1&0  \\
$P_{N-1}:=A^{N-1}(A \tilde{Q})^3$&1&\scriptsize $(N-4)$-th symm.&${\tiny \overline{ \yng(1)}}$&$N+2$&3&0  \\   \hline
$Y_{SU(2N-1)}^{bare}$ & \tiny \begin{tabular}{c}$U(1)_2$ charge: \\ $-2(2N-1)(2N-5)$　\end{tabular}&1&1&$2-4N$&$-4$&2  \\
 \small  $Y^{d}:=Y_{SU(2N-1)}^{bare} A^{2N-5}$ &1&\scriptsize $(2N-5)$-th symm.&1&$-2N-3$&$-4$&2  \\
 \hline
  \end{tabular}}
  \end{center}\label{SU(2N+1)2anti04}
\end{table}

\subsection{$SU(2N+1)$ with $\protect\Young[-0.5]{11}+2 \, \protect\Young[0]{1}+(2N+1) \, \overline{\protect\Young[0]{1}}$}
Let us move on to the $SU(2N+1)$ gauge theory with a single anti-symmetric matter. The UV description is a 3d $\mathcal{N}=2$ $SU(2N+1)$ gauge theory with an anti-symmetric matter, two fundamental matters and $2N+1$ anti-fundamental matters. This theory was studied in \cite{Nii:2016jzi} by using the de-confinement method. Since the theory is ``chiral,'' the bare Coulomb branch $Y^{bare}_{SU(2N-1)}$ is not gauge-invariant. The $U(1)_2$ charge of $Y^{bare}_{SU(2N-1)}$ is $2(2N-1)$ which is positive and this can be made neutral by using $ {\tiny \yng(1,1)}_{\, 0,-4} \in A$ and ${\tiny \yng(1)}_{\, 0,-2} \in Q$. The dressed operator is defined by
\begin{align}
Y^{dressed}_{AQ}:= Y^{bare}_{SU(2N-1)} \left( {\tiny \yng(1,1)}_{\, 0,-4} \right)^{N-1} {\tiny \yng(1)}_{\, 0,-2}\sim Y^{bare}_{SU(2N-1)} A^{N-1} Q.
\end{align}
Note that the dressed operator has a flavor index of $Q$. The elementary fields and their quantum numbers are summarized in Table \ref{SU(2N+1)2}.
The low-energy dynamics is described by the gauge singlet chiral superfields in Table \ref{SU(2N+1)2} and the superpotential
\begin{align}
W=Y^{dressed}_{AQ} \left[ \bar{B}_1^{N} M +B_{N} \bar{B}  \right].
\end{align}
\begin{table}[H]\caption{3d $\mathcal{N}=2$ $SU(2N+1)$ with ${\tiny \protect\yng(1,1)} +2\, {\tiny \protect\yng(1)}+ (2N+1) \,{\tiny \overline{\protect\yng(1)}}$} 
\begin{center}
\scalebox{0.76}{
  \begin{tabular}{|c||c||c|c|c|c|c|c| } \hline
  &$SU(2N+1)$&$SU(2)$&$SU(2N+1)$&$U(1)$&$U(1)$&$U(1)$&$U(1)_R$  \\ \hline
 $A$ &${\tiny \yng(1,1)}$&1&1&1&0&0&$0$ \\
 $Q$ & ${\tiny \yng(1)}$ & ${\tiny \yng(1)}$&1&0&1&0&$0$ \\
$\tilde{Q}$  &${\tiny \overline{\yng(1)}}$&1&${\tiny \yng(1)}$&0&$0$&1&$0$ \\  \hline
$M:=Q\tilde{Q}$&1&${\tiny \yng(1)}$&${\tiny \yng(1)}$&0&1&1&$0$ \\ 
$B_{N}:=A^{N}Q$&1& ${\tiny \yng(1)}$&1&$N$&1&0&$0$  \\
$\bar{B}:= \tilde{Q}^{2N+1}$&1&1&1&$0$&0&$2N+1$&$0$  \\
$\bar{B}_1:=A \tilde{Q}^{2}$&1&1&${\tiny \yng(1,1)}$&1&0&2&$0$  \\  \hline
$Y^{bare}_{SU(2N-1)}$&\scriptsize $U(1)_2$ charge: $2(2N-1)$&1&1&$-(2N-1)$&$-2$&$-(2N+1)$&$2$  \\
$Y^{dressed}_{AQ}:=Y^{bare}_{SU(2N-1)} A^{N-1} Q$&1&${\tiny \yng(1)}$&1&$-N$&$-1$&$-(2N+1)$&$2$  \\
 \hline
  \end{tabular}}
  \end{center}\label{SU(2N+1)2}
\end{table}

\subsection{$SU(2N+1)$ with $\protect\Young[-0.5]{11}+3 \, \protect\Young[0]{1}+2N \, \overline{\protect\Young[0]{1}}$}
The next example is a 3d $\mathcal{N}=2$ $SU(2N+1)$ gauge theory with an anti-symmetric matter, three fundamental matters and $2N$ anti-fundamental matters. The elementary fields and their quantum numbers are summarized in Table \ref{SU(2N+1)3}. Since this theory is ``vector-like'' in a sense that the corresponding 4d theory has no chiral gauge anomaly, the bare Coulomb branch operator $Y^{bare}_{SU(2N-1)}$ is gauge-invariant and becomes a moduli coordinate. The low-energy effective theory is described by the gauge invariant operators in Table \ref{SU(2N+1)3} and the superpotential
\begin{align}
W= Y \left[ \bar{B}_1^N B_{N-1} +\bar{B}_1^{N-1} M^2 B_N \right]
\end{align}
\begin{table}[H]\caption{3d $\mathcal{N}=2$ $SU(2N+1)$ with ${\tiny \protect\yng(1,1)} +3\, {\tiny \protect\yng(1)}+2N \,{\tiny \overline{\protect\yng(1)}}$} 
\begin{center}
\scalebox{0.9}{
  \begin{tabular}{|c||c||c|c|c|c|c|c| } \hline
  &$SU(2N+1)$&$SU(3)$&$SU(2N)$&$U(1)$&$U(1)$&$U(1)$&$U(1)_R$  \\ \hline
 $A$ &${\tiny \yng(1,1)}$&1&1&1&0&0&$0$ \\
 $Q$ & ${\tiny \yng(1)}$ & ${\tiny \yng(1)}$&1&0&1&0&$0$ \\
$\tilde{Q}$  &${\tiny \overline{\yng(1)}}$&1&${\tiny \yng(1)}$&0&$0$&1&$0$ \\  \hline
$M:=Q\tilde{Q}$&1&${\tiny \yng(1)}$&${\tiny \yng(1)}$&0&1&1&$0$ \\ 
$B_{N}:=A^{N}Q$&1& ${\tiny \yng(1)}$&1&$N$&1&0&$0$  \\
$B_{N-1}:=A^{N-1}Q^3$&1& $1$&1&$N-1$&3&0&$0$  \\
$\bar{B}_1:=A \tilde{Q}^{2}$&1&1&${\tiny \yng(1,1)}$&1&0&2&$0$  \\  \hline
$Y^{bare}_{SU(2N-1)}$&1&1&1&$-(2N-1)$&$-3$&$-2N$&$2$  \\
 \hline
  \end{tabular}}
  \end{center}\label{SU(2N+1)3}
\end{table}

\subsection{$SU(2N+1)$ with $\protect\Young[-0.5]{11}+4 \, \protect\Young[0]{1}+(2N-1) \, \overline{\protect\Young[0]{1}}$}
The UV description is a 3d $\mathcal{N}=2$ $SU(2N+1)$ gauge theory with an anti-symmetric matter, four fundamental matters and $2N-1$ anti-fundamental matters. In this example, the bare Coulomb branch operator $Y^{bare}_{SU(2N-1)}$ obtains a non-zero $U(1)_2$ charge. This operator can be made neutral by defining the dressed operators
\begin{align}
Y^{dressed}_{A} &:=Y^{bare}_{SU(2N-1)}  \mathbf{1}_{0,4N-2} \sim Y^{bare}_{SU(2N-1)} A \\
Y^{dressed}_{\tilde{Q}} &:= Y^{bare}_{SU(2N-1)} \left(  {\tiny \overline{\yng(1)}}_{\, 0,2} \right)^{2N-1} \sim Y^{bare}_{SU(2N-1)} \tilde{Q}^{2N-1},
\end{align}
where the color indices of $\tilde{Q}$ are contracted by the epsilon tensor of the $SU(2N-1)$ gauge group. Table \ref{SU(2N+1)4} summarizes the quantum numbers of the matter fields and the moduli coordinates. The confining superpotential can be determined from the symmetry argument as
\begin{align}
W= Y^{dressed}_{A} \left[ \bar{B}_1^{N-1}M B_{N-1} +\bar{B}_1^{N-2} M^3 B_N  \right] +Y^{dressed}_{\tilde{Q}} B_N B_{N-1}.
\end{align}

\begin{table}[H]\caption{3d $\mathcal{N}=2$ $SU(2N+1)$ with ${\tiny \protect\yng(1,1)} +4\, {\tiny \protect\yng(1)}+ (2N-1) \,{\tiny \overline{\protect\yng(1)}}$} 
\begin{center}
\scalebox{0.75}{
  \begin{tabular}{|c||c||c|c|c|c|c|c| } \hline
  &$SU(2N+1)$&$SU(4)$&$SU(2N-1)$&$U(1)$&$U(1)$&$U(1)$&$U(1)_R$  \\ \hline
 $A$ &${\tiny \yng(1,1)}$&1&1&1&0&0&$0$ \\
 $Q$ & ${\tiny \yng(1)}$ & ${\tiny \yng(1)}$&1&0&1&0&$0$ \\
$\tilde{Q}$  &${\tiny \overline{\yng(1)}}$&1&${\tiny \yng(1)}$&0&$0$&1&$0$ \\  \hline
$M:=Q\tilde{Q}$&1&${\tiny \yng(1)}$&${\tiny \yng(1)}$&0&1&1&$0$ \\ 
$B_{N}:=A^{N}Q$&1& ${\tiny \yng(1)}$&1&$N$&1&0&$0$  \\
$B_{N-1}:=A^{N-1}Q^3$&1& ${\tiny \overline{\yng(1)}}$&1&$N-1$&3&0&$0$  \\
$\bar{B}_1:=A \tilde{Q}^{2}$&1&1&${\tiny \yng(1,1)}$&1&0&2&$0$  \\  \hline
$Y^{bare}_{SU(2N-1)}$&\scriptsize $U(1)_2$ charge: $-2(2N-1)$&1&1&$-(2N-1)$&$-4$&$-(2N-1)$&$2$  \\
$Y^{dressed}_{A}:=Y^{bare}_{SU(2N-1)} A$&1&1&1&$-(2N-2)$&$-1$&$-(2N-1)$&$2$  \\
$Y^{dressed}_{\tilde{Q}}:=Y^{bare}_{SU(2N-1)} \tilde{Q}^{2N-1}$&1&1&1&$-(2N-2)$&$-4$&$0$&$2 $  \\
 \hline
  \end{tabular}}
  \end{center}\label{SU(2N+1)4}
\end{table}

\subsection{$SU(2N+1)$ with $\protect\Young[-0.5]{11}+\overline{\protect\Young[-0.5]{11}}+2 ( \protect\Young[0]{1}+ \overline{\protect\Young[0]{1}} )$}
Next, we move on to the $SU(2N+1)$ gauge theories with an anti-symmetric flavor. The first example is a 3d $\mathcal{N}=2$ $SU(2N+1)$ gauge theory with an anti-symmetric flavor and two (anti-)fundamental flavors. This example was studied in \cite{Nii:2016jzi, Csaki:2014cwa}. Table \ref{SU(2N+1)antiflavor22} summarizes the quantum numbers of the elementary fields and the moduli coordinates. The theory is ``vector-like'' in a sense that the corresponding 4d theory has no gauge anomaly. Therefore, the Coulomb branch operator $Y^{bare}_{SU(2N-1)}$ is gauge-invariant. Notice that the anti-symmetric matter reduces two massless components along the Coulomb branch labeled by $Y^{bare}_{SU(2N-1)}$ 
\begin{align}
{\tiny \yng(1,1)} \ni {\tiny \yng(1,1)}_{\, 0,-4}+\mathbf{1}_{0,4N-2},~~~~{\tiny \overline{\yng(1,1)}} \ni {\tiny \overline{\yng(1,1)}}_{\, 0,4} +\mathbf{1}_{0,-(4N-2)}.
\end{align}
As a result, along the Coulomb branch, there are two types of gauge invariants
\begin{align}
\left( {\tiny \yng(1,1)}_{\, 0,-4} {\tiny \overline{\yng(1,1)}}_{\, 0,4} \right)^{a},~~~\left( \mathbf{1}_{0,4N-2} \mathbf{1}_{0,-(4N-2)}  \right)^a.
\end{align}
These are not represented by the operator products between $Y^{bare}_{SU(2N-1)}$ and $T_n$.
This results in the following dressed Coulomb branch operators
\begin{align}
Y_{a}:= Y^{bare}_{SU(2N-1)}  \left( \mathbf{1}_{0,4N-2} \mathbf{1}_{0,-(4N-2)}  \right)^a \sim Y^{bare}_{SU(2N-1)} (A \tilde{A})^a,~~~~a=0,\cdots,N-1.
\end{align}
These operators should be regarded as the moduli coordinates independent of $Y^{bare}_{SU(2N-1)} T_a$. The low-energy dynamics is described by the gauge-invariant fields in Table \ref{SU(2N+1)antiflavor22}. The confining superpotential can be determined for each $N$.

\begin{table}[H]\caption{3d $\mathcal{N}=2$ $SU(2N+1)$ with $\protect\Young[-0.5]{11}+\overline{\protect\Young[-0.5]{11}}+2 ( \protect\Young[0]{1}+ \overline{\protect\Young[0]{1}} )$} 
\begin{center}
\scalebox{0.78}{
  \begin{tabular}{|c||c||c|c|c|c|c|c|c| } \hline
  &$SU(2N+1)$&$SU(2)$&$SU(2)$&$U(1)$&$U(1)$&$U(1)$&$U(1)$&$U(1)_R$  \\ \hline
 $A$ &${\tiny \yng(1,1)}$&1&1&1&0&0&0&$0$ \\
  $\tilde{A}$ &${\tiny \overline{\yng(1,1)}}$&1&1&0&1&0&0&$0$ \\
 $Q$ & ${\tiny \yng(1)}$ & ${\tiny \yng(1)}$&1&0&0&1&0&$0$ \\
$\tilde{Q}$  &${\tiny \overline{\yng(1)}}$&1&${\tiny \yng(1)}$&0&$0$&0&1&$0$ \\  \hline
$M_{k=0,\cdots,N-1}:=Q(A\tilde{A})^k \tilde{Q}$&1&${\tiny \yng(1)}$&${\tiny \yng(1)}$&$k$&$k$&1&1&0  \\
$H_{k=0,\cdots,N-1}:=\tilde{A}(A \tilde{A})^k Q^2$&1&1&1&$k$&$k+1$&2&0&0  \\
$\bar{H}_{k=0,\cdots,N-1}=A(A \tilde{A})^k \tilde{Q}^2$&1&1&1&$k+1$&$k$&0&2&0  \\
$B_N:=A^N Q$&1& ${\tiny \yng(1)}$&1&$N$&0&1&0&0  \\
$\bar{B}_{N}:=\tilde{A}^N \tilde{Q}$&1&1& ${\tiny \yng(1)}$&0&$N$&0&1&0  \\
$T_{n=1,\cdots,N}:=(A \tilde{A})^n$&1&1&1&$n$&$n$&0&0&0  \\  \hline
$Y_{a=0,\cdots,N-1}:=Y^{bare}_{SU(2N-1)} (A \tilde{A})^a$&1&1&1&$1-2N+a$&$1-2N+a$&$-2$&$-2$&2  \\
 \hline
  \end{tabular}}
  \end{center}\label{SU(2N+1)antiflavor22}
\end{table}

\subsection{$SU(2N+1)$ with $\protect\Young[-0.5]{11}+\overline{\protect\Young[-0.5]{11}}+3\,  \protect\Young[0]{1}+ \overline{\protect\Young[0]{1}}$}
The final example in this section is a 3d $\mathcal{N}=2$ $SU(2N+1)$ gauge theory with an anti-symmetric flavor, three fundamental matters and one anti-fundamental matter. The elementary fields and their quantum numbers are defined in Table \ref{SU(2N+1)antiflavor31}. Since the theory is ``chiral,'' the bare Coulomb branch operator $Y_{SU(2N-1)}^{bare}$ has a non-zero $U(1)_2$ charge. The dressed Coulomb branch operators are defined by
\begin{align}
Y^{dressed}_{a}&:=Y_{SU(2N-1)}^{bare} \mathbf{1}_{0,4N-2} \left( \mathbf{1}_{0,4N-2} \mathbf{1}_{0,-(4N-2)}  \right)^a  \nonumber \\
& \sim Y_{SU(2N-1)}^{bare} A(A \tilde{A})^a,~~~~~~~~~(a=0,\cdots,N-1) \\
Y_{\tilde{A}}^{dressed}&:= Y_{SU(2N-1)}^{bare} \left( {\tiny \overline{\yng(1,1)}}_{\, 0,4} \right)^{N-1} {\tiny \overline{\yng(1)}}_{\, 0,2} \sim Y_{SU(2N-1)}^{bare}\tilde{A}^{N-1} \tilde{Q},
\end{align}
where the color indices of $\,{\tiny \overline{\yng(1,1)}}_{\, 0,4}$ and $\,{\tiny \overline{\yng(1)}}_{\, 0,2}$ are contracted by the epsilon tensor of the $SU(2N-1)$ gauge group. Notice that a dressed operator such as 
\begin{align}
Y_{SU(2N-1)}^{bare} \left( {\tiny \overline{\yng(1,1)}}_{\, 0,4} \right)^{N-1} {\tiny \overline{\yng(1)}}_{\, 0,2}  \left( \mathbf{1}_{0,4N-2} \mathbf{1}_{0,-(4N-2)}  \right)
\end{align}
is identified with $Y^{dressed}_{a=0} \bar{B}_{N}$ and cannot be an independent operator.
The low-energy effective description is dual to a non-gauge theory with the gauge-invariant chiral superfields in Table \ref{SU(2N+1)antiflavor31}. 

\begin{table}[H]\caption{3d $\mathcal{N}=2$ $SU(2N+1)$ with $\protect\Young[-0.5]{11}+\overline{\protect\Young[-0.5]{11}}+3\,  \protect\Young[0]{1}+ \overline{\protect\Young[0]{1}}$} 
\begin{center}
\scalebox{0.78}{
  \begin{tabular}{|c||c||c|c|c|c|c|c| } \hline
  &$SU(2N+1)$&$SU(3)$&$U(1)$&$U(1)$&$U(1)$&$U(1)$&$U(1)_R$  \\ \hline
 $A$ &${\tiny \yng(1,1)}$&1&1&0&0&0&$0$ \\
  $\tilde{A}$ &${\tiny \overline{\yng(1,1)}}$&1&0&1&0&0&$0$ \\
 $Q$ & ${\tiny \yng(1)}$ & ${\tiny \yng(1)}$&0&0&1&0&$0$ \\
$\tilde{Q}$  &${\tiny \overline{\yng(1)}}$&1&0&$0$&0&1&$0$ \\  \hline
$M_{k=0,\cdots,N-1}:=Q(A\tilde{A})^k \tilde{Q}$&1&${\tiny \yng(1)}$&$k$&$k$&1&1&0  \\
$H_{k=0,\cdots,N-2}:=\tilde{A}(A \tilde{A})^k Q^2$&1&${\tiny \overline{\yng(1)}}$&$k$&$k+1$&2&0&0  \\
$B_N:=A^NQ$&1& ${\tiny \overline{\yng(1)}}$&$N$&0&1&0&0  \\
$\bar{B}_{N}:=\tilde{A}^N \tilde{Q}$&1&1&0&$N$&0&1&0  \\
$B_{N-1}:=A^{N-1}Q^3$&1&1&$N-1$&0&3&0&0  \\
$T_{n=1,\cdots,N}:=(A \tilde{A})^n$&1&1&$n$&$n$&0&0&0  \\  \hline
$Y_{SU(2N-1)}^{bare}$&\scriptsize $U(1)_2$ charge:$-2(2N-1)$&1&$1-2N$&$1-2N$&$-3$&$-1$&2  \\
$Y^{dressed}_{a=0,\cdots,N-1}:=Y_{SU(2N-1)}^{bare} A(A \tilde{A})^a$&1&1&$2-2N+a$&$1-2N+a$&$-3$&$-1$&2  \\
$Y_{\tilde{A}}^{dressed}:=Y_{SU(2N-1)}^{bare}\tilde{A}^{N-1} \tilde{Q}$&1&1&$1-2N$&$-N$&$-3$&$0$&2  \\
 \hline
  \end{tabular}}
  \end{center}\label{SU(2N+1)antiflavor31}
\end{table}

\section{$SU(4)$ gauge theories}
In this section, we study the s-confinement phases of the 3d $\mathcal{N}=2$ $SU(4)$ gauge theories with anti-symmetric matters. Some examples were already studied in Section 2 while others are new and include three anti-symmetric tensors. In all the examples, we will compute the superconformal indices and find a perfect agreement.  

\subsection{$SU(4)$ with $3 \, \protect\Young[-0.5]{11}+ \protect\Young[0]{1}+  \overline{\protect\Young[0]{1}}$}
The first example is a 3d $\mathcal{N}=2$ $SU(4)$ gauge theory with three anti-symmetric matters and a single flavor in (anti-)fundamental representations. This theory is equivalent to the 3d $\mathcal{N}=2$ $Spin(6)$ gauge theory with three vectors and two spinors. The Coulomb branch of the $Spin(N)$ theory was studied in \cite{Aharony:2011ci, Aharony:2013kma, Nii:2018tnd, Nii:2018wwj}. The corresponding 4d theory was studied in \cite{Grinstein:1997zv, Grinstein:1998bu} and we can derive the 4d result from a 3d perspective.
  
Let us investigate the Coulomb branch. In this example, we need to introduce a different Coulomb branch. The Coulomb branch denoted by $Y_{SU(2) \times SU(2)}$ corresponds to the gauge symmetry breaking
\begin{align}
SU(4) & \rightarrow SU(2) \times SU(2) \times U(1) \nonumber \\
{\tiny \yng(1,1)} &  \rightarrow ({\tiny \yng(1)},{\tiny \yng(1)})_0 +(1,1)_2 +(1,1)_{-2}   \\
{\tiny \yng(1)} & \rightarrow  ({\tiny \yng(1)},0)_1+(1,{\tiny \yng(1)})_{-1} \\
{\tiny \overline{\yng(1)}}  & \rightarrow ({\tiny \yng(1)},0)_{-1}+(1,{\tiny \yng(1)})_{1},
\end{align}
where all the components of (anti-)fundamental matters become massive and integrated out from the low-energy spectrum. The anti-symmetric matter reduces to a massless $({\tiny \yng(1)},{\tiny \yng(1)})_0$ which makes the vacuum of the low-energy $SU(2) \times SU(2)$ gauge theory stable and supersymmetric.  
On the other hand, the Coulomb branch $Y_{SU(2)}$ associated with the gauge symmetry breaking
\begin{align}
SU(4) & \rightarrow SU(2) \times U(1)_1 \times U(1)_2  \nonumber \\
{\tiny \yng(1,1)} &  \rightarrow  {\tiny \yng(1)}_{1,0}+{\tiny \yng(1)}_{-1,0}+1_{0,2}+1_{0,-2}  \\
{\tiny \yng(1)} & \rightarrow {\tiny \yng(1)}_{0,-1}+1_{1,1}+1_{-1,1} \\
{\tiny \overline{\yng(1)}}  & \rightarrow {\tiny \yng(1)}_{0,1}+1_{-1,-1}+1_{1,-1}
\end{align}
is quantum-mechanically not allowed since the low-energy $SU(2)$ gauge theory only has two fundamental matters and the origin of the moduli space of the low-energy theory is excluded \cite{Aharony:1997bx, deBoer:1997kr}. The low-energy dynamics is described by the gauge singlet fields defined in Table \ref{SU(4)311} and a confining superpotential

\begin{align}
W= Y_{SU(2) \times SU(2)}  \left[ T^3 M_0^2 +TM_2^2+P_3 \bar{P}_{3}\right] + \eta Y_{SU(2) \times SU(2)}M_0,
\end{align}
where the last term is a non-perturbative superpotential generated by a KK-monopole which is a twisted instanton in the corresponding 4d theory on a circle. By integrating out the Coulomb branch operator, which corresponds to the 4d limit, we reproduce the quantum-deformed moduli space \cite{Grinstein:1997zv, Grinstein:1998bu}.  

\begin{table}[H]\caption{3d $\mathcal{N}=2$ $SU(4)$ with ${\tiny 3 \, \protect\yng(1,1)} + {\tiny \protect\yng(1)}+ \,{\tiny \overline{\protect\yng(1)}}$} 
\begin{center}
\scalebox{1}{
  \begin{tabular}{|c||c||c|c|c|c|c| } \hline
  &$SU(4)$&$SU(3)$&$U(1)$&$U(1)$&$U(1)$&$U(1)_R$  \\ \hline
 $A$ &${\tiny \yng(1,1)}$&${\tiny \yng(1)}$ &1&0&0&$0$ \\
 $Q$ & ${\tiny \yng(1)}$ &1&0&1&0&$0$ \\
$\tilde{Q}$  &${\tiny \overline{\yng(1)}}$&1&0&$0$&1&$0$ \\  \hline
$M_0:=Q\tilde{Q}$&1&1&0&1&1&0 \\  
$M_2:=QA^2\tilde{Q}$&1&${\tiny \overline{\yng(1)}}$&2&1&1&$0$  \\
$T:=A^2$&1& ${\tiny \yng(2)}$&2&0&0&$0$  \\
$P_{3}:=QA^3Q$&1&1&3&2&0&$0$ \\
$\bar{P}_{3}:= \tilde{Q}A^3  \tilde{Q}$&1&1&3&0&2&$0$ \\ \hline
$Y_{SU(2) \times SU(2)}$&1&1&$-6$&$-2$&$-2$&$2 $  \\  \hline
  \end{tabular}}
  \end{center}\label{SU(4)311}
\end{table}

As a further consistency check of our low-energy analysis, we can compute the superconformal indices from the electric (UV) and dual (confining) descriptions by using the localization technique \cite{Bhattacharya:2008bja, Kim:2009wb, Imamura:2011su, Kapustin:2011jm}. We find that these two theories give the identical superconformal indices

\scriptsize
\begin{align}
I &=1+x^{1/3} \left(\frac{1}{t^2 u^2 v^6}+t u+6 v^2\right)+x^{2/3} \left(\frac{1}{t^4 u^4 v^{12}}+\frac{6}{t^2 u^2 v^4}+t^2 u^2+\frac{1}{t u v^6}+9 t u v^2+21 v^4\right)+x^{5/6} \left(t^2 v^3+u^2 v^3\right) \nonumber \\
&\qquad +x \left(\frac{1}{t^6 u^6 v^{18}}+\frac{6}{t^4 u^4 v^{10}}+\frac{1}{t^3 u^3 v^{12}}+t^3 u^3+9 t^2 u^2 v^2+\frac{21}{t^2 u^2 v^2}+39 t u v^4+\frac{9}{t u v^4}+56 v^6+\frac{1}{v^6}\right) \nonumber \\
&\qquad \qquad +x^{7/6} \left(t^3 u v^3+6 t^2 v^5+t u^3 v^3+6 u^2 v^5\right) +\cdots,
\end{align}
\normalsize
where $t, u$ and $v$ are the fugacities for the $U(1)$ symmetries of $A,Q$ and $\tilde{Q}$. The r-charges of the elementary fields are fixed to be $r_A=r_Q =r_{\tilde{Q}}  =1/6$. The second term $x^{1/3} \left(\frac{1}{t^2 u^2 v^6}+t u+6 v^2\right)$ is identified with a sum of three operators $Y_{SU(2) \times SU(2)} +M_0+T$. The meson $M_2$ is represented as $3tuv^2 x^{2/3}$. $P_3$ and $\bar{P}_3$ corresponds to the fourth term $x^{5/6} \left(t^2 v^3+u^2 v^3\right)$. From the superconformal index calculation, we can see a non-zero contribution from the sector with a GNO charge $(1,0,0,-1)$ which formally corresponds to the Coulomb branch $Y_{SU(2)}^{bare}:~SU(4) \rightarrow SU(2) \times U(1) \times U(1)$. However, this state must be regarded as a operator product $Y_{SU(2) \times SU(2)} M_0$ since we cannot turn on $M_0$ onto the state with a GNO charge $(1,1,-1,-1)$, where the (anti-)fundamental quarks are all massive. In this way, the lower orders of the superconformal indices can be interpreted as symmetric products between $Y_{SU(2) \times SU(2)}$ and the Higgs branch operators. This is consistent with our analysis of the Coulomb branch.

\subsection{$SU(4)$ with $3 \, \protect\Young[-0.5]{11}+ 2 \, \protect\Young[0]{1}$}
The UV description is a 3d $\mathcal{N}=2$ $SU(4)$ gauge theory with three anti-symmetric matters and two fundamental matters. The analysis of the Coulomb branch is the same as the previous example. The Coulomb branch $Y_{SU(2) \times SU(2)}$ can be stable due to the massless components $({\tiny \yng(1)},{\tiny \yng(1)})_0$. The Coulomb branch $Y_{SU(2)}$ cannot be exactly massless since the low-energy $SU(2)$ theory only has two massless doublets. The quantum numbers of the elementary fields and the moduli coordinates are summarized in Table \ref{SU(4)320}. The confining superpotential is easily written down as follows.
\begin{align}
W= Y_{SU(2) \times SU(2)} \left[ T^2 B^2 +P_3^2  \right]
\end{align}
The F-flatness condition for $Y_{SU(2) \times SU(2)}$ imposes one constraint on the Higgs branch operators and the total number of the Higgs branch coordinates reduces to $11$ which is correctly the classical dimension of the Higgs branch.

\begin{table}[H]\caption{3d $\mathcal{N}=2$ $SU(4)$ with ${\tiny 3 \, \protect\yng(1,1)} +2 \, {\tiny \protect\yng(1)}$} 
\begin{center}
\scalebox{1}{
  \begin{tabular}{|c||c||c|c|c|c|c| } \hline
  &$SU(4)$&$SU(3)$&$SU(2)$&$U(1)$&$U(1)$&$U(1)_R$  \\ \hline
 $A$ &${\tiny \yng(1,1)}$&${\tiny \yng(1)}$ &1&1&0&$0$ \\
 $Q$ & ${\tiny \yng(1)}$ &1&${\tiny \yng(1)}$&0&1&$0$ \\ \hline
$T:=A^2$&1& ${\tiny \yng(2)}$&1&2&0&$0$  \\
$B:=AQ^2$&1&${\tiny \yng(1)}$&1&1&2&$0$  \\
$P_{3}:=QA^3Q$&1&1&${\tiny \yng(2)}$&3&2&$0$ \\ \hline
$Y_{SU(2) \times SU(2)}$&1&1&$1$&$-6$&$-4$&$2$  \\  \hline
  \end{tabular}}
  \end{center}\label{SU(4)320}
\end{table}

As a check of our analysis, we compute the superconformal indices by using the electric and dual (confining) descriptions. These two descriptions give us an identical result for the superconformal indices
\small
\begin{align}
I &=1+6 v^2 x^{1/4}+3 t^2 v x^{3/8}+21 v^4 x^{1/2}+21 t^2 v^3 x^{5/8}+x^{3/4} \left(\frac{1}{t^4 v^6}+6 t^4 v^2+56 v^6\right)\nonumber \\
&\quad +81 t^2 v^5 x^{7/8} +x \left(45 t^4 v^4+\frac{6}{t^4 v^4}+126 v^8\right)+x^{9/8} \left(10 t^6 v^3+231 t^2 v^7+\frac{3}{t^2 v^5}\right) \nonumber \\
& \quad +x^{5/4} \left(185 t^4 v^6+\frac{21}{t^4 v^2}+252 v^{10}\right)+x^{11/8} \left(78 t^6 v^5+546 t^2 v^9+\frac{18}{t^2 v^3}\right) \nonumber \\
&\quad +x^{3/2} \left(\frac{1}{t^8 v^{12}}+15 t^8 v^4+555 t^4 v^8+\frac{56}{t^4}+462 v^{12}+\frac{6}{v^4}\right)+x^{13/8} \left(333 t^6 v^7+1134 t^2 v^{11}+\frac{60}{t^2 v}\right) \nonumber \\
&\qquad +x^{7/4} \left(\frac{6}{t^8 v^{10}}+120 t^8 v^6+1365 t^4 v^{10}+\frac{126 v^2}{t^4}+792 v^{14}+\frac{30}{v^2}\right)+\cdots,
\end{align}
\normalsize

\noindent where $t$ and $v$ are the fugacities for the $U(1)$ symmetries of $Q$ and $A$, respectively. The first term $6 v^2 x^{1/4}$ corresponds to $T$. The third term $3 t^2 v x^{3/8}$ is identified with the baryon $B$. $P_3$ is represented as $3 t^2 v^3 x^{5/8}$. The Coulomb branch operator $Y_{SU(2) \times SU(2)}$ corresponds to $\frac{x^{3/4}}{t^4 v^6}$. The lowest order of the indices with a GNO charge $(1,0,0,-1)$ is formally identified with a dressed operator $Y_{SU(2)}^{dress}:= Y_{SU(2)}^{bare} A$. However, this can be regarded as a product $Y_{SU(2) \times SU(2)} B$. In this way, our analysis of the  Coulomb branch is consistent with the superconformal indices.

\subsection{$SU(4)$ with $2 \, \protect\Young[-0.5]{11}+2( \protect\Young[0]{1}+  \overline{\protect\Young[0]{1}})$}
The next example is a 3d $\mathcal{N}=2$ $SU(4)$ gauge theory with two anti-symmetric tensors and two (anti-)fundamental flavors, which was studied in \cite{Csaki:2014cwa, Nii:2016jzi}. Since the theory is now ``vector-like,'' the bare Coulomb branch operators $Y_{SU(2)}^{bare}$ and $Y^{bare}_{SU(2) \times SU(2)}$ are gauge invariant. These Coulomb branches are made stable and supersymmetric due to the various massless components.
The Coulomb branch $Y^{bare}_{SU(2) \times SU(2)}$ can be also interpreted as a dressed operator of $Y_{SU(2)}^{bare}$
\begin{align}
Y^{bare}_{SU(2) \times SU(2)} :=  Y_{SU(2)}^{bare} 1_{0,2} 1_{0,-2} \sim  Y_{SU(2)}^{bare} A^2.
\end{align}
Notice that the product $1_{0,2} 1_{0,-2}$ has four components from the flavor indices of $A^2$ and that the three components are identified with $Y_{SU(2)}^{bare}T$ and a remaining one corresponds to the additional Coulomb branch $Y^{bare}_{SU(2) \times SU(2)}$. The matter fields and their quantum numbers are summarized in Table \ref{SU(4)222}. The confining superpotential takes the following form
\begin{align}
W= Y^{bare}_{SU(2)} \left[ T^2 \det M_0+  \det M_2 + TB \bar{B} \right] +Y^{bare}_{SU(2) \times SU(2)} \left[ M_0 M_2 +B \bar{B} \right],
\end{align}
which is consistent with all the symmetries in Table \ref{SU(4)222}. 

\begin{table}[H]\caption{3d $\mathcal{N}=2$ $SU(4)$ with ${\tiny 2 \, \protect\yng(1,1)} +2 ( {\tiny \protect\yng(1)}+ {\tiny \overline{\protect\yng(1)}})$} 
\begin{center}
\scalebox{1}{
  \begin{tabular}{|c||c||c|c|c|c|c|c|c| } \hline
  &$SU(4)$&$SU(2)$&$SU(2)$&$SU(2)$&$U(1)$&$U(1)$&$U(1)$&$U(1)_R$  \\ \hline
 $A$ &${\tiny \yng(1,1)}$&${\tiny \yng(1)}$ &1&1&1&0&0&$0$ \\
 $Q$ & ${\tiny \yng(1)}$ &1&${\tiny \yng(1)}$&1&0&1&0&$0$ \\ 
$\tilde{Q}$ &${\tiny \overline{\yng(1)}}$&1&1&${\tiny \yng(1)}$&0&0&1&$0$ \\  \hline
 $M_0:=Q \tilde{Q}$&1&1&${\tiny \yng(1)}$&${\tiny \yng(1)}$&0&1&1&$0$  \\
 $M_2:=Q A^2  \tilde{Q}$&1&1&${\tiny \yng(1)}$&${\tiny \yng(1)}$&2&1&1&$0$  \\
 $B:=AQ^2$&1&${\tiny \yng(1)}$&1&1&1&2&0&$0$ \\
 $\bar{B}:=A \tilde{Q}^2$&1&${\tiny \yng(1)}$&1&1&1&0&2&0 \\
$T:=A^2$ &1&${\tiny \yng(2)}$&1&1&2&0&0&$0$ \\  \hline
$Y_{SU(2)}^{bare}$ &1&1&1&1&$-4$&$-2$&$-2$&$2$  \\
 $Y^{bare}_{SU(2) \times SU(2)}$&1&1&1&1&$-2$&$-2$&$-2$&$2$  \\ \hline
  \end{tabular}}
  \end{center}\label{SU(4)222}
\end{table}

We can test the s-confinement phase by computing the superconformal indices. Both the electric and dual (confinement) descriptions give an identical result

\scriptsize
\begin{align}
I&=1+x^{1/4} \left(4 t u+3 v^2\right)+x^{3/8} \left(2 t^2 v+2 u^2 v\right)+x^{1/2} \left(10 t^2 u^2+16 t u v^2+6 v^4\right) \nonumber \\
&\quad+x^{5/8} \left(8 t^3 u v+6 t^2 v^3+8 t u^3 v+6 u^2 v^3\right)  +x^{3/4} \left(3 t^4 v^2+20 t^3 u^3+49 t^2 u^2 v^2+36 t u v^4+3 u^4 v^2+10 v^6\right) \nonumber \\
&\quad +x^{7/8} \left(20 t^4 u^2 v+32 t^3 u v^3+20 t^2 u^4 v+12 t^2 v^5+32 t u^3 v^3+12 u^2 v^5\right) \nonumber \\
&\quad +x \left(12 t^5 u v^2+35 t^4 u^4+9 t^4 v^4+112 t^3 u^3 v^2+126 t^2 u^2 v^4+\frac{1}{t^2 u^2 v^4}+12 t u^5 v^2+64 t u v^6+9 u^4 v^4+15 v^8\right) \nonumber \\
&\quad +x^{9/8} \left(4 t^6 v^3+40 t^5 u^3 v+96 t^4 u^2 v^3+40 t^3 u^5 v+72 t^3 u v^5+96 t^2 u^4 v^3+20 t^2 v^7+72 t u^3 v^5+4 u^6 v^3+20 u^2 v^7\right) \nonumber \\
&\quad +x^{5/4} \left(30 t^6 u^2 v^2+56 t^5 u^5+48 t^5 u v^4+215 t^4 u^4 v^2+18 t^4 v^6+324 t^3 u^3 v^4+30 t^2 u^6 v^2+241 t^2 u^2 v^6+  \right. \nonumber \\
&\qquad   \left.\frac{4 \left(t u+v^2\right)}{t^2 u^2 v^4}+48 t u^5 v^4+100 t u v^8+18 u^4 v^6+21 v^{10}\right)+\cdots,
\end{align}
\normalsize

\noindent  where $t, u$ and $v$ are the fugacities for the global $U(1)$ symmetries for $Q, \tilde{Q}$ and $A$, respectively. The r-charges of the elementary fields are fixed to be $r_A=r_Q=r_{\tilde{Q}} =1/8$ for simplicity. The second term $x^{1/4} \left(4 t u+3 v^2\right)$ consists of $M_0+T$. The third term $x^{3/8} \left(2 t^2 v+2 u^2 v\right)$ is identified with the (anti-)baryons $B$ and $\bar{B}$. $M_2$ is represented as $4tuv^2 x^{1/2}$. The two Coulomb branch operators $Y_{SU(2)}^{bare}$ and $Y^{bare}_{SU(2) \times SU(2)}$ are represented as $\frac{x}{t^2 u^2 v^4}$ and $\frac{x^{5/4}}{t^2 u^2 v^2}$, respectively. 

\subsection{$SU(4)$ with $2 \, \protect\Young[-0.5]{11}+3\, \protect\Young[0]{1}+  \overline{\protect\Young[0]{1}}$}
Let us consider the 3d $\mathcal{N}=2$ $SU(4)$ gauge theory with two anti-symmetric tensors, three fundamental matters and a single anti-fundamental matter. The quantum numbers of the matter fields are summarized in Table \ref{SU(4)231}. Since the theory is ``chiral,'' the bare Coulomb branch operator $Y_{SU(2)}^{bare}$ is not gauge-invariant. In order to cancel the $U(1)_2$ charge of $Y_{SU(2)}^{bare}$, we need to define a dressed operator
\begin{align}
Y_{SU(2)}^{dressed}:=Y_{SU(2)}^{bare} 1_{0,2} \sim Y_{SU(2)}^{bare} A.
\end{align}
Notice that the operator $Y_{SU(2)}^{bare} 1_{0,2} (1_{0,2} 1_{0,-2})$ only has six components from the flavor indices of $A^3$ and these are identified with $Y_{SU(2)}^{dressed} T$. Therefore, the Coulomb branch is described only by $Y_{SU(2)}^{dressed}$ as opposed to the previous subsection where the Coulomb branch dressed by $1_{0,2} 1_{0,-2}$ has to be regarded as an independent operator. 
The low-energy dynamics is dual to a non-gauge theory of the gauge-invariant chiral superfields in Table \ref{SU(4)231} with the superpotential
\begin{align}
W= Y_{SU(2)}^{dressed} \left[ M_2 B +TM_0B \right].
\end{align}

\begin{table}[H]\caption{3d $\mathcal{N}=2$ $SU(4)$ with ${\tiny 2 \, \protect\yng(1,1)} +3 \, {\tiny \protect\yng(1)}+ {\tiny \overline{\protect\yng(1)}}$} 
\begin{center}
\scalebox{0.9}{
  \begin{tabular}{|c||c||c|c|c|c|c|c| } \hline
  &$SU(4)$&$SU(2)$&$SU(3)$&$U(1)$&$U(1)$&$U(1)$&$U(1)_R$  \\ \hline
 $A$ &${\tiny \yng(1,1)}$&${\tiny \yng(1)}$ &1&1&0&0&$r_A$ \\
 $Q$ & ${\tiny \yng(1)}$ &1&${\tiny \yng(1)}$&0&1&0&$r$ \\ 
 $\tilde{Q}$&${\tiny \overline{\yng(1)}}$&1&1&0&0&1&$\bar{r}$ \\  \hline
 $M_0:=Q \tilde{Q}$&1&1& ${\tiny \yng(1)}$&0&1&1&$r+\bar{r}$  \\
 $M_2:=Q A^2\tilde{Q}$&1&1& ${\tiny \yng(1)}$&2&1&1&$2r_A +r +\bar{r}$  \\
  $T:=A^2$&1& ${\tiny \yng(2)}$&1&2&0&0&$2r_A$  \\
   $B:=AQ^2$&1& ${\tiny \yng(1)}$&${\tiny \overline{\yng(1)}}$&1&2&0&$r_A +2r$  \\ \hline
   $Y_{SU(2)}^{bare}$ & \footnotesize $U(1)_2$ charge: $-2$&1&1&$-4$&$-3$&$-1$&$2-4r_A -3r -\bar{r}$   \\
    $Y_{SU(2)}^{dressed}:=Y_{SU(2)}^{bare} A$&1&${\tiny \yng(1)}$&1&$-3$&$-3$&$-1$&$2-3r_A -3r -\bar{r}$ \\  \hline
  \end{tabular}}
  \end{center}\label{SU(4)231}
\end{table}
We can test the validity of our analysis by computing the superconformal indices. The electric and dual descriptions give us the following indices
\small
\begin{align}
 I &= 1+\frac{2 x^{1/4}}{t^3 u v^3}+3x^{1/2}\left(\frac{1}{t^6 u^2 v^6}+t u+v^2\right)+x^{3/4} \left(\frac{4}{t^9 u^3 v^9}+\frac{6}{t^3 u v}+\frac{6}{t^2 v^3}+6 t^2 v\right)  \nonumber \\
 & \qquad+x \left(\frac{5}{t^{12} u^4 v^{12}}+\frac{9}{t^6 u^2 v^4}+\frac{9}{t^5 u v^6}+6 t^2 u^2+12 t u v^2+\frac{9}{t u v^2}+6 v^4\right)+\cdots,
\end{align}
\normalsize

\noindent where the r-charges of the elementary fields are set to be $r_A=r_Q = r_{\tilde{Q}}=1/4.$ $t,u$ and $v$ are the fugacities for the $U(1)$ symmetries of $Q, \tilde{Q}$ and $A$, respectively. The second term $\frac{2 x^{1/4}}{t^3 u v^3}$ is the dressed Coulomb branch $Y_{SU(2)}^{dressed}$. The third term includes $M_0$ and $T$. $B$ is represented as $6t^2 v x^{3/4}$ and $M_2$ is $3tuv^2 x$.

\subsection{$SU(4)$ with $2 \, \protect\Young[-0.5]{11}+ 4 \, \protect\Young[0]{1}$}
The final example in this section is a 3d $\mathcal{N}=2$ $SU(4)$ gauge theory with two anti-symmetric tensors and four fundamental matters. Table \ref{SU(4)240} summarizes the elementary fields and their quantum numbers. The bare Coulomb branch operator $Y_{SU(2)}^{bare}$ is not gauge-invariant and the dressed operator is defined by
\begin{align}
Y_{SU(2)}^{dressed} :=  Y_{SU(2)}^{bare} (1_{0,2})^2 \sim Y_{SU(2)}^{bare}  A^2,
\end{align}
where the flavor indices of $A^2$ are totally symmetrized. 
The low-energy effective theory is described by the gauge singlets listed in Table \ref{SU(4)240} and a confining superpotential
\begin{align}
W= Y_{SU(2)}^{dressed} (TB+B_1^2).
\end{align}
Notice that $Y_{SU(2)}^{bare} (1_{0,2})^2 (1_{0,2} 1_{0,-2})$ has eight components from the flavor indices of $A^4$ and these are identified with a product $Y_{SU(2)}^{dressed} T$. This is consistent with our analysis of the Coulomb branch and the above superpotential since the F-flatness condition for $B$ imposes one constraint on $Y_{SU(2)}^{dressed} T$ which has nine components.  

\begin{table}[H]\caption{3d $\mathcal{N}=2$ $SU(4)$ with ${\tiny 2 \, \protect\yng(1,1)} + 4 \, {\tiny \protect\yng(1)}$} 
\begin{center}
\scalebox{1}{
  \begin{tabular}{|c||c||c|c|c|c|c| } \hline
  &$SU(4)$&$SU(2)$&$SU(4)$&$U(1)$&$U(1)$&$U(1)_R$  \\ \hline
 $A$ &${\tiny \yng(1,1)}$&${\tiny \yng(1)}$ &1&1&0&$0$ \\
 $Q$ & ${\tiny \yng(1)}$ &1&${\tiny \yng(1)}$&0&1&$0$ \\ \hline
$T:=A^2$&1& ${\tiny \yng(2)}$&1&2&0&$0$  \\
$B:=Q^4$&1&1&1&1&4&$0$  \\
$B_1:=AQ^2$&1&${\tiny \yng(1)}$&${\tiny \yng(1,1)}$&1&2&$0$  \\ \hline
$Y_{SU(2)}^{bare} $& \footnotesize $U(1)_2$ charge: $-4$&1&$1$&$-4$&$-4$&$2  $  \\  
$Y_{SU(2)}^{dressed} :=Y_{SU(2)}^{bare}  A^2$&1&${\tiny \yng(2)}$&1&$-2$&$-4$&$2$   \\ \hline
  \end{tabular}}
  \end{center}\label{SU(4)240}
\end{table}

Let us test the validity of the s-confinement phase by computing the superconformal indices. The electric (UV) theory and the dual (confinement) description give an identical result:
\small
\begin{align}
I &=1+x^{1/2} \left(\frac{3}{t^4 v^2}+3 v^2\right)+12 t^2 v x^{3/4}+x \left(\frac{6}{t^8 v^4}+t^4+\frac{8}{t^4}+6 v^4\right)+x^{5/4} \left(36 t^2 v^3+\frac{24}{t^2 v}\right) \nonumber \\ 
&\qquad +x^{3/2} \left(\frac{10}{t^{12} v^6}+\frac{15}{t^8 v^2}+78 t^4 v^2+\frac{15 v^2}{t^4}+10 v^6\right)+x^{7/4} \left(\frac{36}{t^6 v^3}+12 t^6 v+72 t^2 v^5+\frac{60 v}{t^2}\right) \nonumber \\
&\qquad \quad +x^2 \left(\frac{15}{t^{16} v^8}+\frac{24}{t^{12} v^4}+t^8+\frac{27}{t^8}+231 t^4 v^4+\frac{24 v^4}{t^4}-\frac{3}{t^4 v^4}+15 v^8+80\right)+\cdots,
\end{align}
\normalsize

\noindent where the r-charges of the elementary fields are set to be $r_A=r_Q =1/4.$ $t$ and $v$ are the fugacities for the $U(1)$ global symmetries rotating $Q$ and $A$, respectively. The second term $x^{1/2} \left(\frac{3}{t^4 v^2}+3 v^2\right)$ is regarded as a sum $Y_{SU(2)}^{dressed}+T$. The third term $12 t^2 v x^{3/4}$ corresponds to $B_1$. The operator $B$ is represented as $t^4 x$. The higher order terms are the symmetric products of these moduli operators and the fermion contributions.

\section{$SU(5)$ gauge theories}
In this section, we will investigate the s-confinement phases in the 3d $\mathcal{N}=2$ $SU(5)$ gauge theories with anti-symmetric matters. Some examples were already studied in the previous section while other examples are new and include three anti-symmetric tensors.

\subsection{$SU(5)$ with $2 \, \protect\Young[-0.5]{11}+2( \protect\Young[0]{1}+ \overline{\protect\Young[0]{1}})$}
The first example is a 3d $\mathcal{N}=2$ $SU(5)$ with two anti-symmetric matters and two (anti-)fundamental flavors. This case was studied in Section 3. The bare Coulomb branch corresponds to the breaking $SU(5) \rightarrow SU(3) \times U(1)_1 \times U(1)_2$ and the dressed operator is defined by
\begin{align}
Y_{SU(3)}^{dressed}:=  Y_{SU(3)}^{bare}  \mathbf{1}_{0,6}  \sim Y_{SU(3)}^{bare}A
\end{align}
The low-energy dynamics is described by the moduli coordinates defined in Table \ref{SU(5)222} and a confining superpotential
\begin{align}
W=  Y_{SU(3)}^{dressed} \left[M B_2 \bar{B}_3 +B_2^2 \bar{B}_1 +\bar{B}_3 P  \right].
\end{align}

\begin{table}[H]\caption{3d $\mathcal{N}=2$ $SU(5)$ with ${\tiny 2 \, \protect\yng(1,1)} + {\tiny 2\,( \protect\yng(1)}+ {\tiny \overline{\protect\yng(1)}})$} 
\begin{center}
\scalebox{0.9}{
  \begin{tabular}{|c||c||c|c|c|c|c|c|c| } \hline
  &$SU(5)$&$SU(2)$&$SU(2)$&$SU(2)$&$U(1)$&$U(1)$&$U(1)$&$U(1)_R$  \\ \hline
 $A$ &${\tiny \yng(1,1)}$&${\tiny \yng(1)}$ &1&1&1&0&0&$0$ \\
 $Q$ & ${\tiny \yng(1)}$ &1&${\tiny \yng(1)}$&1&0&1&0&$0$ \\
 $\tilde{Q}$ & ${\tiny \overline{ \yng(1)}}$ &1&1&${\tiny \yng(1)}$&0&0&1&$0$ \\  \hline
 $M:=Q \tilde{Q}$&1&1&${\tiny \yng(1)}$&${\tiny \yng(1)}$&0&1&1&$0$  \\
 $B_2:=A^2 Q$&1&${\tiny \yng(2)}$&${\tiny \yng(1)}$&1&2&1&0&$0$  \\
 $\bar{B}_1:=A \tilde{Q}^2$&1&${\tiny \yng(1)}$&1&1&1&0&2&$0$  \\
 $\bar{B}_3:=A^3 \tilde{Q}$&1&${\tiny \yng(1)}$&1&${\tiny \yng(1)}$&3&0&1&$0$  \\
 $P:=A^2Q^2 \tilde{Q}$&1&1&1&${\tiny \yng(1)}$&2&2&1&$0$ \\  \hline 
 $Y_{SU(3)}^{bare}$&\footnotesize $U(1)_2$ charge: $-6$&1&1&1&$-6$&$-2$&$-2$&$2$ \\
 $Y_{SU(3)}^{dressed}:= Y_{SU(3)}^{bare}A$&1&${\tiny \yng(1)}$&1&1&$-5$&$-2$&$-2$& $2$ \\ \hline
  \end{tabular}}
  \end{center}\label{SU(5)222}
\end{table}

\subsection{$SU(5)$ with $2 \, \protect\Young[-0.5]{11}+ \protect\Young[0]{1}+ 3 \,  \overline{\protect\Young[0]{1}}$}
The UV description is a 3d $\mathcal{N}=2$ $SU(5)$ gauge theory with two anti-symmetric tensors, one fundamental matter and three anti-fundamental matters. Table \ref{SU(5)213} summarizes the quantum numbers of the elementary fields and the moduli coordinates. Since the theory is ``vector-like,'' the bare Coulomb branch operator $Y_{SU(3)}^{bare}$ is gauge invariant. In this theory, we can also turn on another Coulomb branch $Y_{SU(2) \times SU(2)}^{bare}$ whose vev leads to the breaking
\begin{align}
SU(5) & \rightarrow SU(2) \times SU(2) \times U(1)_1 \times U(1)_2 \\
{\tiny \yng(1)} & \rightarrow ({\tiny \yng(1)},1)_{1,1}+(1,{\tiny \yng(1)})_{-1,1}+(1,1)_{0,-4} \\
 {\tiny \overline{ \yng(1)}} & \rightarrow  ({\tiny \yng(1)},1)_{-1,-1}+(1,{\tiny \yng(1)})_{1,-1}+(1,1)_{0,4} \\
 {\tiny \yng(1,1)} & \rightarrow ({\tiny \yng(1)},{\tiny \yng(1)})_{0,2}+({\tiny \yng(1)},1)_{1,-3}+(1,{\tiny \yng(1)})_{-1,-3}+(1,1)_{2,2}+(1,1)_{-2,2}.
\end{align}
Since the bare operator $Y_{SU(2) \times SU(2)}^{bare}$ has a non-zero $U(1)_2$ charge proportional to the mixed Chern-Simons term between $U(1)_1$ and $U(1)_2$, the dressed operator is defined by
\begin{align}
Y_{SU(2) \times SU(2)}^{dressed}:= Y_{SU(2) \times SU(2)}^{bare} (1,1)_{0,-4} \sim Y_{SU(2) \times SU(2)}^{bare} Q.
\end{align}
Notice that this dressed operator can be interpreted as a dressed operator of $Y_{SU(3)}^{bare}$ as follows
\begin{align}
Y_{SU(2) \times SU(2)}^{dressed} \sim Y_{SU(3)}^{bare} \mathbf{1}_{0,6} {\tiny \overline{ \yng(1)}}_{\, 0,-4}  {\tiny \yng(1)}_{\, 0,-2} \sim  Y_{SU(3)}^{bare}  A^2 Q , 
\end{align}
where the right hand side has four components from the flavor indices of $A^2$. The three components are identified with $ Y_{SU(3)}^{bare} B_2$ and a remaining component is identified with $Y_{SU(2) \times SU(2)}^{dressed}$. By using the moduli coordinates listed in Table \ref{SU(5)213}, we find an s-confinement phase with the superpotential
\begin{align}
W= Y_{SU(3)}^{bare} \left[ M \bar{B}_3^2 +B_2 \bar{B}_1 \bar{B}_3 \right] +Y^{dressed}_{SU(2) \times SU(2)} \left[ \bar{B}_1 \bar{B}_3 \right].
\end{align}

When the corresponding 4d theory is put on a circle, the twisted instanton generates a non-perturbative superpotential $\Delta W = \eta Y_{SU(3)}^{bare}$. By integrating out the two Coulomb branch, we can reproduce the 4d result \cite{Aharony:1997bx}. In a 4d limit, we obtain two constraints
\begin{align}
M \bar{B}_3^2 +B_2 \bar{B}_1 \bar{B}_3 +\eta =0,~~~~ \bar{B}_1 \bar{B}_3=0
\end{align}
These are consistent with the 4d quantum-deformed moduli space \cite{Grinstein:1997zv}.

\begin{table}[H]\caption{3d $\mathcal{N}=2$ $SU(5)$ with ${\tiny 2 \, \protect\yng(1,1)} + {\tiny \protect\yng(1)}+3 \, {\tiny \overline{\protect\yng(1)}}$} 
\begin{center}
\scalebox{0.9}{
  \begin{tabular}{|c||c||c|c|c|c|c|c| } \hline
  &$SU(5)$&$SU(2)$&$SU(3)$&$U(1)$&$U(1)$&$U(1)$&$U(1)_R$  \\ \hline
 $A$ &${\tiny \yng(1,1)}$&${\tiny \yng(1)}$ &1&1&0&0&$0$ \\
 $Q$ & ${\tiny \yng(1)}$ &1&1&0&1&0&$0$ \\
 $\tilde{Q}$ & ${\tiny \overline{ \yng(1)}}$ &1&${\tiny \yng(1)}$&0&0&1&$0$ \\  \hline
$M:=Q \tilde{Q}$&1&1& ${\tiny \yng(1)}$ &0&1&1&$0$ \\
 $B_2 :=A^2 Q$&1& ${\tiny \yng(2)}$ &1&2&1&0&$0$  \\
 $\bar{B}_1:=A \tilde{Q}^2$&1& ${\tiny \yng(1)}$ & ${\tiny \overline{\yng(1)}}$ &1&0&2&$0$  \\
 $\bar{B}_3:=A^2 (A \tilde{Q})$&1& ${\tiny \yng(1)}$ & ${\tiny \yng(1)}$ &3&0&1&$0$  \\ \hline
 $Y_{SU(3)}^{bare}$&1&1&1&$-6$&$-1$&$-3$&$2$ \\
 $Y_{SU(2) \times SU(2)}^{bare}$&\footnotesize $U(1)_2$ charge: $4$&1&1&$-4$&$-1$&$-3$&$2$ \\
  $Y_{SU(2) \times SU(2)}^{dressed}:=Y_{SU(2) \times SU(2)}^{bare} Q$&1&1&1&$-4$&$0$&$-3$&$2$ \\ \hline
  \end{tabular}}
  \end{center}\label{SU(5)213}
\end{table}

\subsection{$SU(5)$ with $2 \, \protect\Young[-0.5]{11}+3\, \protect\Young[0]{1}+ \overline{\protect\Young[0]{1}}$}
The next example is a 3d $\mathcal{N}=2$ $SU(5)$ gauge theory with two anti-symmetric tensors, three fundamental matters and one anti-fundamental matter. The matter fields and their quantum numbers are summarized in Table \ref{SU(5)231}. Since the theory is ``chiral,'' the bare Coulomb branch $Y_{SU(3)}^{bare}$ is not gauge-invariant. The dressed (gauge-invariant) operator is defined by 
\begin{align}
Y_{SU(3)}^{dressed}:=Y_{SU(3)}^{bare} \mathbf{1}_{0,6}^2 \sim Y_{SU(3)}^{bare}A^2,
\end{align}
where the flavor indices of $A^2$ are symmetrized.
The low-energy effective theory is described by the gauge-invariant operators in Table \ref{SU(5)231} and a confining superpotential
\begin{align}
W= Y_{SU(3)}^{dressed} \left[ M B_2^2 +B_1 \bar{B}_3 +B_2 P \right]
\end{align}

\begin{table}[H]\caption{3d $\mathcal{N}=2$ $SU(5)$ with ${\tiny 2 \, \protect\yng(1,1)} + 3 \, {\tiny \protect\yng(1)}+{\tiny \overline{\protect\yng(1)}}$} 
\begin{center}
\scalebox{0.9}{
  \begin{tabular}{|c||c||c|c|c|c|c|c| } \hline
  &$SU(5)$&$SU(2)$&$SU(3)$&$U(1)$&$U(1)$&$U(1)$&$U(1)_R$  \\ \hline
 $A$ &${\tiny \yng(1,1)}$&${\tiny \yng(1)}$ &1&1&0&0&$r_A$ \\
 $Q$ & ${\tiny \yng(1)}$ &1&${\tiny \yng(1)}$&0&1&0&$r$ \\
 $\tilde{Q}$ & ${\tiny \overline{ \yng(1)}}$ &1&1&0&0&1&$\bar{r}$ \\  \hline
$M:=Q \tilde{Q}$&1&1& ${\tiny \yng(1)}$ &0&1&1&$r+\bar{r}$ \\
$B_1 :=A Q^3$&1&${\tiny \yng(1)}$&1&1&3&0&$r_A+3r$ \\
 $B_2 :=A^2 Q$&1& ${\tiny \yng(2)}$ &${\tiny \yng(1)}$&2&1&0&$2r_A +r$  \\
 $\bar{B}_3:=A^2 (A \tilde{Q})$&1& ${\tiny \yng(1)}$ & 1 &3&0&1&$3r_A+\bar{r}$  \\ 
 $P:=A^2 Q^2  \tilde{Q}$&1&1& ${\tiny \overline{ \yng(1)}}$&2&2&1&$2r_A+2r +\bar{r}$  \\ \hline
 $Y_{SU(3)}^{bare}$&\footnotesize $U(1)_2$ charge: $-12$&1&1&$-6$&$-3$&$-1$&$2-6r_A -3r -\bar{r}$ \\
 $Y_{SU(3)}^{dressed}:=Y_{SU(3)}^{bare}A^2$&1&${\tiny \yng(2)}$&1&$-4$&$-3$&$-1$&$2-4r_A -3r -\bar{r}$  \\  \hline
   \end{tabular}}
  \end{center}\label{SU(5)231}
\end{table}

\subsection{$SU(5)$ with $2 \, \protect\Young[-0.5]{11}+4\, \overline{\protect\Young[0]{1}}$}
Let us consider the 3d $\mathcal{N}=2$ $SU(5)$ gauge theory with two anti-symmetric tensors and four anti-fundamental matters. The bare Coulomb branch operator $Y_{SU(3)}^{bare}$ has a positive $U(1)_2$ charge $6$. Therefore, the gauge-invariant dressed operator becomes
\begin{align}
Y_{SU(3)}^{dressed}:=Y_{SU(3)}^{bare} \left( {\tiny \overline{ \yng(1)}}_{0,-4} \right)^2   {\tiny \overline{ \yng(1)}}_{0,2}  \sim Y_{SU(3)}^{bare}A^2 \tilde{Q},
\end{align}
where the flavor indices of $A^2$ are anti-symmetrized. The low-energy dynamics is described by the moduli operators in Table \ref{SU(5)204} and the superpotential 
\begin{align}
W= Y_{SU(3)}^{dressed} \bar{B}_1 \bar{B}_3.
\end{align}

\begin{table}[H]\caption{3d $\mathcal{N}=2$ $SU(5)$ with ${\tiny 2 \, \protect\yng(1,1)} +4 \, {\tiny \overline{\protect\yng(1)}}$} 
\begin{center}
\scalebox{1}{
  \begin{tabular}{|c||c||c|c|c|c|c| } \hline
  &$SU(5)$&$SU(2)$&$SU(4)$&$U(1)$&$U(1)$&$U(1)_R$  \\ \hline
 $A$ &${\tiny \yng(1,1)}$&${\tiny \yng(1)}$ &1&1&0&$0$ \\
 $\tilde{Q}$ & ${\tiny \overline{ \yng(1)}}$ &1&${\tiny \yng(1)}$&0&1&$0$ \\  \hline
 $\bar{B}_1:=A \tilde{Q}$&1&${\tiny \yng(1)}$&${\tiny \yng(1,1)}$&1&2&$0$\\
 $\bar{B}_3:=\tilde{Q}$&1&${\tiny \yng(1)}$&${\tiny \yng(1)}$&3&1&$0$\\ \hline
 $Y_{SU(3)}^{bare}$&\footnotesize $U(1)_2$ charge: $6$&1&1&$-6$&$-4$&$2$\\
 $Y_{SU(3)}^{dressed}:=Y_{SU(3)}^{bare}A^2 \tilde{Q}$ &1&1&${\tiny \yng(1)}$&$-4$&$-3$&$2$  \\ \hline
  \end{tabular}}
  \end{center}\label{SU(5)204}
\end{table}

\subsection{$SU(5)$ with $2 \, \protect\Young[-0.5]{11}+4\, \protect\Young[0]{1}$}
We next study the 3d $\mathcal{N}=2$ $SU(5)$ gauge theory with two anti-symmetric tensors and four fundamental matters. The bare Coulomb branch $Y_{SU(3)}^{bare}$ has a $U(1)_2$ charge $-18$ and then the dressed operator becomes
\begin{align}
Y_{SU(3)}^{dressed}:=Y_{SU(3)}^{bare} \mathbf{1}_{0,6}^3  \sim Y_{SU(3)}^{bare}A^3
\end{align}
Table \ref{SU(5)240} summarizes the quantum numbers of the elementary fields and the moduli coordinates. The low-energy dynamics is described by these moduli fields and the superpotential
\begin{align}
W= Y_{SU(3)}^{dressed} B_1 B_2.
\end{align}
\begin{table}[H]\caption{3d $\mathcal{N}=2$ $SU(5)$ with ${\tiny 2 \, \protect\yng(1,1)} +4 \, {\tiny \protect\yng(1)}$} 
\begin{center}
\scalebox{1}{
  \begin{tabular}{|c||c||c|c|c|c|c| } \hline
  &$SU(5)$&$SU(2)$&$SU(4)$&$U(1)$&$U(1)$&$U(1)_R$  \\ \hline
 $A$ &${\tiny \yng(1,1)}$&${\tiny \yng(1)}$ &1&1&0&$0$ \\
 $Q$ & ${\tiny  \yng(1)}$ &1&${\tiny \yng(1)}$&0&1&$0$ \\  \hline
 $B_1 :=A Q^3$&1&${\tiny \yng(1)}$ &${\tiny \overline{ \yng(1)}}$ &1&3&$0$  \\
 $B_2:=A^2Q$&1&${\tiny \yng(2)}$ &${\tiny \yng(1)}$ &2&1&$0$  \\  \hline 
 $Y_{SU(3)}^{bare}$&\footnotesize $U(1)_2$ charge: $-18$&1&1&$-6$&$-4$&$2$  \\
 $Y_{SU(3)}^{dressed}:=Y_{SU(3)}^{bare}A^3$&1&${\tiny \yng(3)}$&1&$-3$&$-4$&$2$  \\ \hline
   \end{tabular}}
  \end{center}\label{SU(5)240}
\end{table}

\subsection{$SU(5)$ with $3 \, \protect\Young[-0.5]{11} +   \protect\Young[0]{1}$}
Next, we will study the $SU(5)$ gauge theory with three anti-symmetric tensors. The first example is a 3d $\mathcal{N}=2$ $SU(5)$ gauge theory with three anti-symmetric tensors and a single fundamental matter. Notice that the Coulomb branch described by $Y_{SU(3)}^{bare}$ cannot become exactly massless since the low-energy $SU(3)$ gauge theory only includes one fundamental matter and three anti-fundamental matters and since the origin of its vacuum is excluded from the moduli space. We instead have to consider the Coulomb branch $Y^{bare}_{SU(2) \times SU(2)}$ whose vev leads to the gauge symmetry breaking $SU(5) \rightarrow SU(2) \times SU(2) \times U(1)_1 \times U(1)_2$. Since the theory is ``chiral,'' $Y^{bare}_{SU(2) \times SU(2)}$ has a non-zero $U(1)_2$ charge. Therefore, we need to introduce a dressed operator
\begin{align}
Y^{d}:=Y^{bare}_{SU(2) \times SU(2)}  (1,1)_{0,-4} \sim Y^{bare}_{SU(2) \times SU(2)} Q.
\end{align}
The superconformal index calculation naively tells us that there is a contribution from a state with a GNO charge $(1,0,0,-1)$, whose operator form is 
\begin{align}
Y_{SU(3)}^{bare} \mathbf{1}_{0,6}^2  \sim Y_{SU(3)}^{bare}  A^2.
\end{align}
However, this can be identified with $Y^d T$ and cannot be an independent operator. The low-energy dynamics is described by the gauge invariant fields defined in Table \ref{SU(5)3anti1f} and a confining superpotential
\begin{align}
W=Y^d B T^2.
\end{align}

\begin{table}[H]\caption{3d $\mathcal{N}=2$ $SU(5)$ with ${\tiny 3 \, \protect\yng(1,1)} + {\tiny \protect\yng(1)}$} 
\begin{center}
\scalebox{1}{
  \begin{tabular}{|c||c||c|c|c|c| } \hline
  &$SU(5)$&$SU(3)$&$U(1)$&$U(1)$&$U(1)_R$  \\ \hline
 $A$ &${\tiny \yng(1,1)}$&${\tiny \yng(1)}$ &1&0& 0 \\
 $ Q$ & ${\tiny  \yng(1)}$ &1&0&1&0 \\  \hline 
 $B:=A^2 Q $&1& ${\tiny  \yng(2)}$&2&1&0  \\
$T:=A^5$ &1& ${\tiny  \yng(2)}$&5&0&0  \\  \hline
 $Y^{bare}_{SU(2) \times SU(2)}$&$U(1)_2$: $4$&1&$-12$&$-2$&2  \\
 $Y^{d}:=Y^{bare}_{SU(2) \times SU(2)} Q$&1&1&$-12$&$-1$&2  \\  \hline
   \end{tabular}}
  \end{center}\label{SU(5)3anti1f}
\end{table}

\subsection{$SU(5)$ with $3 \, \protect\Young[-0.5]{11} +   \overline{\protect\Young[0]{1}}$}
The next example is a 3d $\mathcal{N}=2$ $SU(5)$ gauge theory with three anti-symmetric tensors and a single anti-fundamental matter. In this theory, the Coulomb branch $Y^{bare}_{SU(3)}$ is allowed since the low-energy $SU(3)$ theory has four anti-fundamental matters and since its vacuum is stable and supersymmetric. However, due to the ``chirality'' of the theory, $Y^{bare}_{SU(3)}$ is not gauge-invariant. The dressed (gauge-invariant) operator is defined by
\begin{align}
Y^{d}:=Y^{bare}_{SU(3)} \mathbf{1}_{0,6} \sim Y^{bare}_{SU(3)} A,
\end{align}
which is fundamental under the global $SU(3)$ symmetry. For the bare Coulomb branch labeled by $Y_{SU(2) \times SU(2)}^{bare}$ which has a non-zero $U(1)_2$ charge, we cannot define a gauge singlet. Therefore, the Coulomb branch is three-dimensional and described by $Y^{d}$. The low-energy dynamics is given by the gauge singlets defined in Table \ref{SU(5)3anti1af} and the superpotential
\begin{align}
W=Y^d P_2 T \label{WSU(5)3anti1af}
\end{align}

This theory is related to the s-confinement phase in a 4d $\mathcal{N}=1$ $SU(5)$ gauge theory with three anti-symmetric tensors and three anti-fundamental matters via real mass deformation \cite{Aharony:2013dha, Aharony:2013kma}. The 4d theory has an enhanced flavor symmetry $SU(3)$ of $\tilde{Q}$. The low-energy dynamics is described by $\bar{B}^{4d}:=A\tilde{Q}^2$, $P_2^{4d} :=A^3 \tilde{Q}$, $T^{4d} =A^5$ and a superpotential \cite{Csaki:1996zb}
\begin{align}
W_{4d} = \bar{B}^{4d} P_2^{4d}T^{4d}  + (P_2^{4d})^3. \label{W4dSU(5)3anti1af}
\end{align}
In order to derive the 3d dynamics, we put the 4d theory on a circle. The theory includes a non-perturbative superpotential from a twisted instanton which is known as a KK-monopole. 
By introducing a real mass to a generator $i \sigma_3$ of the $SU(2)$ subgroup in $SU(3)_{\tilde{Q}}$, this non-perturbative effect is turned-off \cite{Aharony:2013dha, Aharony:2013kma} and we can flow to the 3d theory discussed here. On the dual (confining) side, the components charged under the $SU(2)$ subgroup are all massive and integrated out.  
The 4d confining superpotential \eqref{W4dSU(5)3anti1af} correctly reduces to  \eqref{WSU(5)3anti1af} by identifying the massless components of $\bar{B}^{4d}$ with $Y^d$.

\begin{table}[H]\caption{3d $\mathcal{N}=2$ $SU(5)$ with ${\tiny 3 \, \protect\yng(1,1)} + {\tiny \overline{ \protect\yng(1)}}$} 
\begin{center}
\scalebox{1}{
  \begin{tabular}{|c||c||c|c|c|c| } \hline
  &$SU(5)$&$SU(3)$&$U(1)$&$U(1)$&$U(1)_R$  \\ \hline
 $A$ &${\tiny \yng(1,1)}$&${\tiny \yng(1)}$ &1&0& 0 \\
 $ \tilde{Q}$ & ${\tiny  \yng(1)}$ &1&0&1&0 \\  \hline 
 $P_2:=A^2 (A \tilde{Q}) $&1& ${\tiny  \yng(2,1)}$&3&1&0  \\
$T:=A^5$ &1& ${\tiny  \yng(2)}$&5&0&0  \\  \hline
 $Y^{bare}_{SU(3)}$&$U(1)_2$: $-6$&1&$-9$&$-1$&2  \\
 $Y^{d}:=Y^{bare}_{SU(3)} A$&1&${\tiny  \yng(1)}$ &$-8$&$-1$&2  \\  \hline
   \end{tabular}}
  \end{center}\label{SU(5)3anti1af}
\end{table}

\subsection{$SU(5)$ with $2 \, \protect\Young[-0.5]{11}+\overline{\protect\Young[-0.5]{11}}+    \overline{\protect\Young[0]{1}}$}
The UV description is a 3d $\mathcal{N}=2$ $SU(5)$ gauge theory with two anti-symmetric tensors, an anti-symmetric-bar tensor and an anti-fundamental matter. The corresponding 4d theory was studied in \cite{Grinstein:1997zv}. The quantum numbers of the matter fields and the moduli coordinates are summarized in Table \ref{SU(5)2101}. We point out that the Higgs branch operator $T_7:=A^6 \bar{A}$ was missed in \cite{Grinstein:1997zv}. 

The Coulomb branch $Y_{SU(3)}^{bare}$ is gauge-invariant since the theory is ``vector-like'' in a sense that the corresponding 4d theory has no gauge anomaly. The other Coulomb branch is described by a bare operator $Y_{SU(2) \times SU(2)}^{bare}$ which leads to the gauge symmetry breaking $SU(5) \rightarrow SU(2) \times SU(2) \times U(1)_1 \times U(1)_2$. The $U(1)_2$ charge of $Y_{SU(2) \times SU(2)}^{bare}$ is $4$ and this can be dressed as
\begin{align}
Y_{SU(2) \times SU(2)}^{dressed}:= Y_{SU(2) \times SU(2)}^{bare} \left( ({\tiny \yng(1)},{\tiny \yng(1)})_{0,-2} \right)^2  \sim Y_{SU(2) \times SU(2)}^{bare} \tilde{A}^2
\end{align}
The low-energy dynamics is described by the gauge invariant operators in Table \ref{SU(5)2101} and a confining superpotential. 
\begin{align}
W&=Y_{SU(3)}^{bare} \left[ T_4 P_{4,1,1} +T_2^2 P_{4,1,1} +\bar{B}_2 T_7\right] \nonumber \\
&\qquad + Y_{SU(2) \times SU(2)}^{dressed} \left[ T_4 B_3^2 + P_{4,1,1}T_2 B_3 +(T_2 B_3)^3 +P_{2,1,2} T_7 \right].
\end{align}

When we put the corresponding 4d theory on a circle, there is an additional (twisted) instanton (known as a KK-monopole) which generates a non-perturbative superpotential
\begin{align}
\Delta W= \eta Y_{SU(3)}^{bare}
\end{align}
The 4d result can be obtained by integrating out the Coulomb branch coordinates with $\Delta W$ \cite{Aharony:1997bx, Aharony:2013dha, Aharony:2013kma}. In a 4d limit, there are two constraints
\begin{align}
T_4 P_{4,1,1} +T_2^2 P_{4,1,1} +\bar{B}_2 T_7 + \eta &=0 \\
 T_4 B_3^2 + P_{4,1,1}T_2 B_3 +(T_2 B_3)^3 +P_{2,1,2} T_7 &=0,
\end{align}
where the first constraint is quantum-mechanically deformed.

\begin{table}[H]\caption{3d $\mathcal{N}=2$ $SU(5)$ with $2 \, {\tiny \protect\yng(1,1)} +  {\tiny \overline{\protect\yng(1,1)} }+ {\tiny \overline{\protect\yng(1)}}$} 
\begin{center}
\scalebox{0.9}{
  \begin{tabular}{|c||c||c|c|c|c|c| } \hline
  &$SU(5)$&$SU(2)$&$U(1)$&$U(1)$&$U(1)$&$U(1)_R$  \\ \hline
 $A$ &${\tiny \yng(1,1)}$&${\tiny \yng(1)}$ &1&0&0&$0$ \\[5pt]
 $\tilde{A}$&${\tiny \overline{\yng(1,1)}}$&1&0&1&0&$0$ \\
 $\tilde{Q}$ & ${\tiny \overline{\yng(1)}}$ &1&0&0&1&$0$ \\\hline 
 $T_2 :=A \tilde{A}$&1&${\tiny \yng(1)}$&1&1&0&$0$ \\
 $T_4:=(A \tilde{A})^2$&1&${\tiny \yng(2)}$&2&2&0&$0$  \\
$T_7:=A^6 \tilde{A}$ &1&1&6&1&0&$0$  \\
 $\bar{B}_2:=\tilde{A}^2 \tilde{Q}$&1&1&0&2&1&$0$  \\
 $B_3:=A^2 (A \tilde{Q})$&1&${\tiny \yng(1)}$&3&0&1&$0$  \\
 $P_{2,1,2}:=(A\tilde{Q})^2 \tilde{A}$&1&1&2&1&2&$0$  \\
 $P_{4,1,1}:=A \tilde{A} A^3 \tilde{Q}$&1&${\tiny \yng(2)}$&4&1&1&$0$  \\ \hline
 $Y_{SU(3)}$&1&1&$-6$&$-3$&$-1$&$2$  \\
 $Y_{SU(2) \times SU(2)}^{bare}$&\footnotesize $U(1)_2$ charge: $4$&1&$-8$&$-4$&$-2$&$2$  \\
$Y_{SU(2) \times SU(2)}^{dressed}:=Y_{SU(2) \times SU(2)}^{bare} \tilde{A}^2$&1&1&$-8$&$-2$&$-2$&$2$  \\ \hline
  \end{tabular}}
  \end{center}\label{SU(5)2101}
\end{table}

\subsection{$SU(5)$ with $2 \, \protect\Young[-0.5]{11}+\overline{\protect\Young[-0.5]{11}}+    \protect\Young[0]{1}$}
The final example is a 3d $\mathcal{N}=2$ $SU(5)$ gauge theory with two anti-symmetric tensors, an anti-symmetric-bar tensor and a fundamental matter. Since the corresponding 4d theory has a gauge anomaly, there is no 4d s-confinement of this type.  
Table \ref{SU(5)2110} summarizes the quantum numbers of the elementary fields and the moduli coordinates. 

In this theory, the Coulomb brach $Y_{SU(3)}$ is not allowed since the low-energy $SU(3)$ theory only includes two (anti-)fundamental flavors and since its vacuum is runaway and unstable \cite{Aharony:1997bx}. The Coulomb branch is instead described by the bare Coulomb branch operator  $Y_{SU(2) \times SU(2)}^{bare}$ whose vev induces the breaking $SU(5) \rightarrow SU(2) \times SU(2) \times U(1)_1 \times U(1)_2$. In this breaking, the low-energy $SU(2) \times SU(2)$ theory has enough massless dynamical matters and its vacuum can be stable and supersymmetric. The confinement phase is described by the moduli fields in Table \ref{SU(5)2110} and the superpotential
\begin{align}
W= Y^{bare}_{SU(2) \times SU(2)} (PF+B_3^2(T_2+T_1^2) +B_2B_3R_1 (T_2+T_1^2) +B_2^2(T_2^2+T_2T_1^2)  )
\end{align}

By introducing a non-zero vev to $B_2$ with rank-one, the gauge group is higgsed into $USp(4)$. The theory flows to a 3d $\mathcal{N}=2$ $USp(4)$ gauge theory with two anti-symmetric tensors and two fundamental matters which will also exhibit s-confinement in the next section.  

\begin{table}[H]\caption{3d $\mathcal{N}=2$ $SU(5)$ with $2 \, {\tiny \protect\yng(1,1)} +  {\tiny \overline{\protect\yng(1,1)} }+ {\tiny \protect\yng(1)}$} 
\begin{center}
\scalebox{1}{
  \begin{tabular}{|c||c||c|c|c|c|c| } \hline
  &$SU(5)$&$SU(2)$&$U(1)$&$U(1)$&$U(1)$&$U(1)_R$  \\ \hline
 $A$ &${\tiny \yng(1,1)}$&${\tiny \yng(1)}$ &1&0&0&$0$ \\
 $\tilde{A}$&${\tiny \overline{\yng(1,1)}}$&1&0&1&0&$0$ \\
 $Q$ & ${\tiny \yng(1)}$ &1&0&0&1&$0$ \\\hline 
 $T_1 :=A \tilde{A}$&1&${\tiny \yng(1)}$&1&1&0&$0$ \\
 $T_2:=(A \tilde{A})^2$&1&${\tiny \yng(2)}$&2&2&0&$0$  \\
$F:=A^6 \tilde{A}$ &1&1&6&1&0&$0$  \\
 $B_2:=A^2 Q$&1& ${\tiny \yng(2)}$&2&0&1&$0$  \\
 $B_3:=A^2 (A \tilde{A} Q)$&1&${\tiny \yng(1)}$&3&1&1&$0$  \\
 $P:=A^2\tilde{A}^3 Q^2$&1&1&2&3&2&$0$  \\  \hline
 $Y_{SU(2) \times SU(2)}^{bare}$&1&1&$-8$&$-4$&$-2$&$2$  \\ \hline
  \end{tabular}}
  \end{center}\label{SU(5)2110}
\end{table}

\section{$USp(2N)$ gauge theories}
In this section, we consider the 3d $\mathcal{N}=2$ $USp(2N)$ gauge theories with anti-symmetric and fundamental matters without a tree-level superpotential. These theories were studied in \cite{Benvenuti:2018bav, Amariti:2018wht} (see also \cite{Amariti:2015vwa, Amariti:2015mva, Amariti:2016kat}). When the bare Coulomb branch operator denoted by $Y_{USp(2N-2)}$ obtains a non-zero expectation value, the gauge group is spontaneously broken to 
\begin{align}
USp(2N) & \rightarrow USp(2N-2) \times U(1) \\
{\tiny \yng(1)} &  \rightarrow  {\tiny \yng(1)}_{\, 0}+1_{\pm 1} \\
{\tiny \yng(1,1)} &  \rightarrow  {\tiny \yng(1,1)}_{\, 0}+{\tiny \yng(1)}_{\pm 1} +1_{0}  \\
{\tiny \yng(2)} &  \rightarrow  {\tiny \yng(2)}_{\, 0}+ 1_{0} +{\tiny \yng(1)}_{\pm 1} +1_{\pm 2},
\end{align}
where anti-symmetric representations are traceless. Notice that the anti-symmetric matter reduces to two massless components $ {\tiny \yng(1,1)}_{\, 0}$ and $1_{0}$ except for $N=2$. This fact leads to the following dressed operators
\begin{align}
Y_a := Y_{USp(2N-2)} (1_{0})^{a-1} \sim Y_{USp(2N-2)}  A^{a-1}~~~~(a=1,\cdots,N).
\end{align}
These operators should be regarded as independent operators. When the 3d $\mathcal{N}=2$ $USp(2N)$ gauge theory contains $2F$ fundamental matters and $F_A$ anti-symmetric matters, the monopole configuration associated to $Y_{USp(2N-2)}$ has the fermion zero-modes as in Table \ref{zeromodeUSp}. The adjoint zero-modes come from a gaugino field in the $USp(2N)$ vector multiplet. In the following subsections, we will give a list of the $USp(2N)$ s-confinement. 

\begin{table}[H]\caption{Fermion zero-modes of $Y_{USp(2N-2)}$} 
\begin{center}
  \begin{tabular}{|c||c|c|c| } \hline
  &adjoint& fundamental & anti-symmetric \\ \hline
$Z_{USp(2N-2)}$&$2N$&$2F$&$(2N-2) F_A$  \\  \hline
  \end{tabular}
  \end{center}\label{zeromodeUSp}
\end{table}

\subsection{$USp(2N)$ with $\protect\Young[-0.5]{11}+4\, \protect\Young[0]{1}$}
First, we consider the 3d $\mathcal{N}=2$ $USp(2N)$ gauge theory with an anti-symmetric tensor and four fundamental matters, which was studied in \cite{Benvenuti:2018bav, Amariti:2018wht}. Table \ref{USp(2N)14} summarizes the quantum numbers of the elementary fields and the moduli coordinates. The dressed Coulomb branch $Y_a$ is defined above. The low-energy dynamics is described by $M_k, T_j$ and $Y_a$. 

The confining superpotential can be easily written for each $N$. For $USp(4)$, the confining superpotential becomes 
\begin{align}
W=Y_1 \left[T_2M_0^2 + M_1^2 \right] +Y_2 \left[ M_0 M_1 \right].
\end{align}
For $USp(6)$, the superpotential is determined as
\begin{align}
W &= Y_1 \left[ T_2^2 M_0^2 +T_3 M_0M_1 +T_2 M_0 M_2 +M_2^2  \right]  \nonumber \\
&\qquad +Y_2  \left[ T_3 M_0^2 +M_1M_2 \right] +Y_3 \left[ T_2M_0^2 +M_0 M_2 +M_1^2 \right].
\end{align}

\begin{table}[H]\caption{3d $\mathcal{N}=2$ $USp(2N)$ with ${\tiny  \protect\yng(1,1)} + 4 \, {\tiny \protect\yng(1)}$} 
\begin{center}
\scalebox{0.96}{
  \begin{tabular}{|c||c||c|c|c|c| } \hline
  &$USp(2N)$&$SU(4)$&$U(1)$&$U(1)$&$U(1)_R$  \\ \hline
 $A$ &${\tiny \yng(1,1)}$&1&1&0&$0$ \\
 $Q$ & ${\tiny \yng(1)}$ &${\tiny \yng(1)}$&0&1&$0$ \\  \hline
 \small $M_k:= QA^kQ$  ($k=0, \cdots, N-1$)&1&${\tiny \yng(1,1)}$&$k$&2&$0$ \\
\small $T_j:=A^j$ $(j=2,\cdots,N)$&1&1&$j$&0&$0$  \\ \hline
\small $Y_a:=Y_{USp(2N-2)}  A^{a-1}$ $(a=1,\cdots,N)$&1&1&$-(2N-a-1)$&$-4$&$2$  \\ \hline
  \end{tabular}}
  \end{center}\label{USp(2N)14}
\end{table}

These results are consistent with deformations of the electric and magnetic theories. Let us consider a particular Higgs branch by giving a non-zero vev to the anti-symmetric matter as
\begin{align}
\braket{A} =v \, i \sigma_2 \otimes \begin{pmatrix}
\omega_1 & & \\
 & \ddots & \\
&& \omega_N
\end{pmatrix},
\end{align}
where $\omega_i$ are the $N$-th roots of unity. The gauge group is higgsed into $SU(2)^N$ and the the theory flows to decoupled $N$ copies of 3d $\mathcal{N}=2$ $SU(2)$ gauge theories with four doublets, which again shows s-confinement \cite{Aharony:1997bx, deBoer:1997kr}. The same flow can be obtained on the magnetic side. For $N=2$, a non-zero vev is turned on for $\braket{T_2} =v^2$ and the mesonic operators are decomposed into
\begin{align}
M_0 =N_1+N_2,~~~M_1=v(N_1-N_2).
\end{align}
The superpotential becomes 
\begin{align}
W=(v^2 Y_1+vY_2)N_1^2 +(v^2 Y_1-vY_2)N_2^2  =: Y_{SU(2)_1}N_1^2+ Y_{SU(2)_2}N_2^2,
\end{align}
which is a sum of s-confinement phases of two $SU(2)$ gauge theories with four doublets.

We can also test this s-confinement phase by computing the superconformal indices. The electric (UV) and confining descriptions for $N=2$ lead to the identical indices

\scriptsize
\begin{align}
I&=1+x^{1/4} \left(6 t^2+u^2\right)+6 t^2 u x^{3/8}+x^{1/2} \left(21 t^4+6 t^2 u^2+u^4\right)+x^{5/8} \left(35 t^4 u+6 t^2 u^3\right)+x^{3/4} \left(56 t^6+41 t^4 u^2+6 t^2 u^4+u^6\right) \nonumber \\
&+x^{7/8} \left(120 t^6 u+35 t^4 u^3+6 t^2 u^5\right)+x \left(126 t^8+170 t^6 u^2+41 t^4 u^4+6 t^2 u^6+u^8\right)+x^{9/8} \left(315 t^8 u+170 t^6 u^3+35 t^4 u^5+6 t^2 u^7\right) \nonumber \\
& \quad+x^{5/4} \left(252 t^{10}+510 t^8 u^2+170 t^6 u^4+41 t^4 u^6+\frac{1}{t^4 u^2}+6 t^2 u^8+u^{10}\right) \nonumber \\
&\qquad +x^{11/8} \left(700 t^{10} u+595 t^8 u^3+170 t^6 u^5+35 t^4 u^7+\frac{1}{t^4 u}+6 t^2 u^9\right)+\cdots
\end{align}
\normalsize
\noindent where $t$ and $u$ are the fugacities for the $U(1)$ symmetries of $Q$ and $A$, respectively. The r-charges of the elementary fields are fixed to be $r_A=r_Q=1/8$ for simplicity. The second term $x^{1/4} \left(6 t^2+u^2\right)$ corresponds to $M_0$ and $T_2$. The third term $6 t^2 u x^{3/8}$ corresponds to $M_1$. The Coulomb branch operators $Y_1$ and $Y_2$ are represented as $\frac{x^{5/4}}{t^4 u^2}$ and $\frac{x^{11/8}}{t^4 u}$, respectively.

\subsection{$USp(4)$ with $2 \, \protect\Young[-0.5]{11}+2\, \protect\Young[0]{1}$}
Let us consider the 3d $\mathcal{N}=2$ $USp(4)$ gauge theory with two anti-symmetric tensors and two fundamental matters. This is equivalent to the 3d $\mathcal{N}=2$ $Spin(5)$ gauge theory with two vectors and two spinors. For the Coulomb branch of the $Spin(N)$ theory, see \cite{Aharony:2011ci, Aharony:2013kma, Nii:2018tnd, Nii:2018wwj}.
In this example, the Coulomb branch $Y_{USp(2)}$ is not allowed since the low-energy $USp(2)$ theory only has two fundamental matters and a quantum effect excludes the origin of the moduli space \cite{Aharony:1997bx, deBoer:1997kr}. We have to consider another Coulomb branch $Y_{SO(3)}$ which leads to the gauge symmetry breaking
\begin{align}
USp(4) & \rightarrow SO(3) \times U(1) \\
\mathbf{4} & \rightarrow \mathbf{2}_1 +\mathbf{2}_{-1} \\
 \mathbf{5} & \rightarrow  \mathbf{3}_0 +\mathbf{1}_{2} +\mathbf{1}_{-2}.
\end{align}
This is a different breaking pattern since all the components of the fundamental representation become massive along the $SO(3)$ branch. In this branch, the low-energy $SO(3)$ theory has two massless vectors and its vacuum remains stable and supersymmetric. The low-energy dynamics is described by the gauge invariant chiral superfields defined in Table \ref{USp(4)22} and a confining superpotential
\begin{align}
W=Y_{SO(3)} \left[ T^2M_0^2 +T M_1^2  +M_2^2 \right] +\eta Y_{SO(3)} M_0
\end{align}
where the last term appears when the corresponding 4d theory is put on a circle. $\eta$ is a one-instanton factor of the 4d theory. By integrating out the Coulomb branch, we can flow to the 4d limit and reproduce the quantum-deformed moduli space \cite{Grinstein:1997zv}.

\begin{table}[H]\caption{3d $\mathcal{N}=2$ $USp(4)$ with ${\tiny 2 \, \protect\yng(1,1)} + 2 \, {\tiny \protect\yng(1)}$} 
\begin{center}
\scalebox{1}{
  \begin{tabular}{|c||c||c|c|c|c|c| } \hline
  &$USp(4)$&$SU(2)$&$SU(2)$&$U(1)$&$U(1)$&$U(1)_R$  \\ \hline
 $A$ &${\tiny \yng(1,1)}$&${\tiny \yng(1)}$&1&1&0&$0$ \\
 $Q$ & ${\tiny \yng(1)}$ & 1&${\tiny \yng(1)}$&0&1&$0$ \\  \hline
$M_0:=QQ$ &1&1&1&0&2&$0$ \\
 $M_1:=QAQ$&1&${\tiny \yng(1)}$&1&1&2&$0$ \\
 $M_2:=QA^2Q$&1&1&${\tiny \yng(2)}$&2&2&$0$ \\
$T:=A^2$ &1&${\tiny \yng(2)}$&1&2&0&$0$ \\  \hline 
$Y_{SO(3)}$&1&1&1&$-4$&$-4$&$2$\\ \hline
  \end{tabular}}
  \end{center}\label{USp(4)22}
\end{table}

As a check of our analysis, we compute the superconformal indices from the electric and dual descriptions. For simplicity, we take the r-charges of the elementary fields to be $r_A =r_Q=1/8$. Both of the descriptions give an identical result

\scriptsize
\begin{align}
I &=1+x^{1/4} \left(t^2+3 u^2\right)+2 t^2 u x^{3/8}+x^{1/2} \left(t^4+6 t^2 u^2+6 u^4\right)+2 t^2 u x^{5/8} \left(t^2+3 u^2\right)+x^{3/4} \left(t^6+9 t^4 u^2+15 t^2 u^4+10 u^6\right) \nonumber \\
&+2 t^2 u x^{7/8} \left(t^4+6 t^2 u^2+6 u^4\right)+x \left(t^8+9 t^6 u^2+29 t^4 u^4+\frac{1}{t^4 u^4}+28 t^2 u^6+15 u^8\right) \nonumber \\
&+2 t^2 u x^{9/8} \left(t^6+8 t^4 u^2+15 t^2 u^4+10 u^6\right) +x^{5/4} \left(t^{10}+9 t^8 u^2+38 t^6 u^4+61 t^4 u^6+\frac{3}{t^4 u^2}+45 t^2 u^8+\frac{1}{t^2 u^4}+21 u^{10}\right) \nonumber \\
&+x^{11/8} \left(2 t^{10} u+16 t^8 u^3+52 t^6 u^5+56 t^4 u^7+30 t^2 u^9+\frac{2}{t^2 u^3}\right)\nonumber \\
&+x^{3/2} \left(t^{12}+9 t^{10} u^2+43 t^8 u^4+95 t^6 u^6+105 t^4 u^8+\frac{6}{t^4}+66 t^2 u^{10}+\frac{3}{t^2 u^2}+28 u^{12}+\frac{1}{u^4}\right)+\cdots,
\end{align}
\normalsize
where $t$ and $u$ are the fugacities for the $U(1)$ symmetries counting the numbers of $Q$ and $A$, respectively. The second term $x^{1/4} \left(t^2+3 u^2\right)$ corresponds to $M_0+T$. The third term $2 t^2 u x^{3/8}$ is identified with $M_1$. $M_2$ corresponds to $3t^2 u^2x^{1/2}$. The Coulomb branch operator $Y_{SO(3)}$ is represented as $\frac{x}{t^4 u^4}$. From the index computation, there is a contribution with a GNO charge $(1,0)$ which naively corresponds to $Y_{USp(2)} \sim \frac{x^{5/4}}{t^2 u^4}$. However, this should be interpreted as a product $Y_{SO(3)} M_0$ in this theory. This is highly consistent with our analysis of the Coulomb branch.

\section{Summary and Discussion}

In this paper, we investigated the s-confinement phases in the 3d $\mathcal{N}=2$ $SU(N)$ and $USp(2N)$ gauge theory with anti-symmetric tensors and (anti-)fundamental matters. By defining the dressed Coulomb branch operators, we found the confining descriptions where the low-energy dynamics is described by a non-gauge theory with the Higgs and Coulomb branch operators. Since there is no chiral anomaly for the gauge symmetry in 3d (except for the parity anomaly), we can construct various confinement phases for theories with ``chiral'' matter contents. We argued that the bare Coulomb branch operator is not gauge-invariant in those ``chiral'' theories and that the Coulomb moduli space is described by the so-called dressed Coulomb branch operators. The precise forms of the dressed operators are drastically changed depending on the matter content. For small gauge groups, $SU(4)$ and $SU(5)$, we can consider the s-confinement with three anti-symmetric tensors. We also argued that those theories have a different types of the Coulomb branch. In the case of the $SU(4)$ and $USp(4)$ gauge theories, we computed the superconformal indices by using the electric and dual (s-confinement) description and found a beautiful agreement.

For other matter contents which are not discussed here, we could not find the s-confinement descriptions for several reasons. For instance, some of the Coulomb branch operators cannot have r-charge 2 and a non-perturbative superpotential is not available. In some cases, the dressed Coulomb branch operator has positive matter $U(1)$ charges and we cannot write down an effective superpotential which is necessary to reduce the independent number of Higgs branch operators. 

In this paper, we didn't consider the s-confinement phases of the Chern-Simons-matter theories. For the CS theories with chiral matter contents, the Coulomb branch will become more complicated as noted in \cite{Aharony:2013dha} (see also \cite{Benini:2011mf, Aharony:2014uya, Nii:2018bgf}). Especially, when the $SU(N)$ gauge theory has only odd numbers of (anti-)fundamental matters, the theory must include the half odd integer Chern-Simons level due to the parity anomaly of the gauge symmetry. It would be important to classify the s-confinement phases in the Chern-Simons-matter theories with ``chiral'' matter contents.

This paper focused only on the s-confinement phases with second-order anti-symmetric tensors. Then, it would be interesting to classify confinement phases with three-index matters. The ``vector-like'' theories with three-index matters were studied in 3d \cite{Nii:2018erm} and 4d \cite{Csaki:1996zb}, where the s-confinement descriptions for the ``vector-like'' theories are presented. It is worthwhile studying the low-energy dynamics of the ``chiral'' theories with three-index matters. We here only dealt with the s-confinement phases in the 3d $\mathcal{N}=2$ gauge theories without a tree-level superpotential. It would be important to search for the s-confinement with a tree-level superpotential or a monopole superpotential \cite{Benini:2017dud}. It is also important to search for the s-confinement of the exceptional gauge groups, which was done only for the $G_2$ case in 3d \cite{Nii:2017npz}. We will leave these problems as future directions.

\section*{Acknowledgments}
I would like to thank Antonio Amariti for valuable discussions and for sharing his unpublished notes.
This work is supported by the Swiss National Science Foundation (SNF) under grant number PP00P2\_183718/1.


\bibliographystyle{ieeetr}
\bibliography{sconfinement_ref}

\end{document}